\documentclass[11pt]{article}

\usepackage{graphicx}
\newcommand{\ba}    {\begin{array}}
\newcommand{\ea}    {\end{array}}
\newcommand{\be}    {\begin{equation}}
\newcommand{\ee}    {\end{equation}}
\newcommand{\bea}    {\begin{eqnarray}}
\newcommand{\eea}    {\end{eqnarray}}

\renewcommand{\hat}{\widehat}

\newcommand{\uM}       {\mbox{\boldmath$M$}}

\newcommand{\ubeta}             {\mbox{\boldmath$\beta$}}
\newcommand{\ugamma}            {\mbox{\boldmath$\gamma$}}

\newcommand{\uiota}             {\mbox{\boldmath$\uiota$}}

\newcommand{\ulambda}           {\mbox{\boldmath$\lambda$}}

\newcommand{\mY}            {\mathring{Y}}
\newcommand{\mX}            {\mathring{X}}
\newcommand{\mU}            {\mathring{u}}
\newcommand{\mbeta}            {\mathring{\beta}}
\newcommand{\mepsilon}            {\mathring{\epsilon}}
\newcommand{\kronecker}{\raisebox{1pt}{\ensuremath{\:\otimes\:}}} 

\newcommand{\zero}{\mbox{\boldmath$0$}}
\newcommand{\indsim}{\sim  \hspace{-17.5pt}\raisebox{6.5pt}{\text{ \tiny{ind}\normalsize}}\ }
\newcommand{\iidsim}{\sim  \hspace{-18.2pt}\raisebox{6.5pt}{\text{ \tiny{i.i.d}\normalsize}}\ }
\newcommand{\given}{\,|\,}

\makeatletter
\newcommand*{\rom}[1]{\expandafter\@slowromancap\romannumeral #1@}
\makeatother

\usepackage{mathtools} 

\usepackage{amsmath,amssymb,latexsym,epsfig,graphics}
\usepackage{natbib,url}
\usepackage{graphicx, color}
\definecolor{brickred}{rgb}{0.6,0,0}
\definecolor{myblue}{rgb}{0.03137255, 0.11372549, 0.34509804}
\definecolor{zhencolor}{rgb}{0.0039 ,   0.3235,    0.2490}
\RequirePackage[colorlinks]{hyperref} \hypersetup{
    colorlinks=true,       
    linkcolor=red,          
    citecolor=myblue,        
    filecolor=magenta,      
    urlcolor=brickred           
}
\usepackage[all]{xy}

\usepackage{amsmath,amssymb,latexsym,epsfig,graphics}

\newcommand{\bds}{\mathcal{B}}
\newcommand{\cen}{\mathcal{G}}
\newcommand{\cl}{\mathcal{C}}

\newcommand{\norm}{\mbox{Normal}}
\newcommand{\igamma}{\mbox{Inverse-Gamma}}
\newcommand{\unif}{\mbox{Uniform}}

\def\boxit#1{\vbox{\hrule\hbox{\vrule\kern6pt
 \vbox{\kern6pt#1\kern6pt}\kern6pt\vrule}\hrule}}

\title{\bf Spatial Clustering of Curves with Functional Covariates: A Bayesian Partitioning Model with Application to Spectra Radiance in Climate Study}
\author{Zhen Zhang$^{1*}$, Chae Young Lim$^2$, Tapabrata Maiti$^{3}$ and Seiji Kato$^4$\\
$^1$ Department of Statistics, University of Chicago, Chicago, Illinois, USA\\
$^2$ Department of Statistics, Seoul National University, Seoul, South Korea\\
$^3$ Department of Statistics \& Probability \\ Michigan State University, East Lansing, Michigan, USA\\ 
$^4$ Science Directorate, NASA Langley Research Center, Hampton, Virginia, USA\\
{$^*$Email:} zhangz19@galton.uchicago.edu
}
\date{} 

\begin{document}
\maketitle

\begin{abstract}
In climate change study, the infrared spectral signatures of climate change have recently been conceptually adopted, and widely applied to identifying and attributing atmospheric composition change.
We propose a Bayesian hierarchical model for spatial clustering of the high-dimensional functional data based on the effects of functional covariates and local features. 
We couple the functional mixed-effects model with a generalized spatial partitioning method for:
(1) producing spatially contiguous clusters for the high-dimensional spatio-functional data; 
(2) improving the computational efficiency via parallel computing over subregions or multi-level partitions; 
and
(3) capturing the near-boundary ambiguity and uncertainty for data-driven partitions. 
We propose a generalized partitioning method which puts less constraints on the shape of spatial clusters. 
Dimension reduction in the parameter space
is also achieved via Bayesian wavelets to alleviate the increasing model complexity introduced by clusters. 
The model 
well captures the regional effects of the atmospheric and cloud properties on the spectral radiance measurements. 
The results elaborate the importance of exploiting spatially contiguous partitions for identifying regional effects and small-scale variability.

\vspace{.3cm}{\bf Keywords:} Spatial clustering; Bayesian wavelets;  Voronoi Tessellation; Functional covariates; High-dimensional data; Parallel computing; Spectral radiance.
\end{abstract}

\section{Introduction}
\label{section:intro}

\subsection{Spectral radiance change studies}
\label{subsec:spec}

The infrared spectral signatures of  climate change have been conceptually adopted, and widely applied to identifying and attributing atmospheric composition change. The spectrally resolved radiances of earth thermal emission spectra have proven to be beneficial for climate change detection, which is essentially a comparison between unforced radiance variability and anthropogenic trend signals.  
Active 
research is ongoing to analyze how the radiances in different spectral bands vary under different controls of geophysical variables such as atmospheric and cloud properties (temperature, tropospheric water vapor, cloud height and fraction, optical thickness and particle size, etc.) to examine whether a detectable trend against natural variability exists \citep[e.g.,][]{hr09, kato11, kato14}.  
The spectral signatures associated with these factors can be uniquely determined by retrieving the cloud properties and atmospheric composition change.
Therefore, it is very important to quantitatively study the spectral radiance data from multiple perspectives which potentially cast light on climate change.  

To understand the effects of small-scale variability on atmospheric temperature, humidity, and cloud property change detection, 
\cite{kato11} derived the cloud fields for monthly $10^{\circ}$ zonal mean spectral radiances from Moderate Resolution Imaging Spectroradiometer (MODIS) spectral radiance observations. 
Two year's worth of MODIS-derived cloud fields from January 2003 through December 2004 are used as a control run. 
Top-of-the-atmosphere (TOA) longwave nadir-view spectra from $50$ to $2760$ cm$^{-1}$ are computed with a $1.0$ cm$^{-1}$ resolution. 
Detailed procedures for computing the spectral radiances are described in \cite{kato11} and the references provided therein. 
The control run hence includes longwave spectra over these $2711$ 
wavenumbers from $50$ to $2760$ cm$^{-1}$ at a total of 18 $10^{\circ}$ latitude zones from $-85^{\circ}$ to $85^{\circ}$ and the $24$ months between 2003 and 2004. 
In addition to the control run, \cite{kato11} independently perturbed $15$ cloud and atmospheric properties for computing the TOA spectral radiance as perturbed runs, with uniform perturbed amounts at all latitudes. 
The $15$ atmospheric and cloud properties (perturbed amount) are:
Temperature near-surface and skin ($+0.2$ K), at surface--200-hPa ($+0.2$ K) and at 200--10-hPa ($-0.2$ K);
Water vapor at surface--500-hPa ($\times 1.025$) and 500--200-hPa ($\times 1.025$); 
Cloud-top height at low-level ($+0.25$ km), midlevel ($+0.25$ km), and high-level ($+0.20$ km); 
Cloud fraction at low-level ($-0.025$), midlevel ($-0.025$), and high-level ($-0.025$); 
Thin ice cloud ($\tau<1$) optical thickness ($\times 1.3$) and Ice cloud particle size ($+1\, \mu$m); 
Water cloud optical thickness ($\times 1.1$) and Water cloud particle size ($+1\, \mu$m). 
\cite{kato11} studied the monthly zonal mean spectral radiance changes due to perturbations by differencing the control run and individual perturbed run spectral radiances, and averaging the zonal mean spectral radiances weighted by their area for the global mean radiance. 
The shapes of the spectral radiance changes from some perturbations, though individually calculated, were closely connected and functionally similar due to the relationship in certain spectral regions  between these perturbed atmospheric properties. 
\cite{kato11} further quantified the relative difference between the spectral radiance change under simultaneous and individual perturbations using the $10^{\circ}$ zonal mean root-mean-square (RMS) over the wavenumber region from $200$ to $2000$ cm$^{-1}$. 
The results suggest that TOA spectral radiance change can be expressed approximately by a linear combination of the radiance changes due to individual perturbation.  
However, the spectral difference is found to be large  
particularly in the presence of the cloud fraction change.
All these facts 
purport the
joint modeling of these perturbed runs to comprehensively study the effects on the TOA radiance or radiance change, which may vary across the spectral regions over wavenumbers. Furthermore, such effects can potentially be 
piecewise linear across the 18 $10^{\circ}$ latitude zones and $24$ months due to the geographical and seasonal impact. A global linearity assumption may 
obscure
such intra-region dissimilarity. 

The earlier study by \cite{kato11} treated the spectral radiance difference between the control runs in 2003 and 2004 as the observed response variable in their linear regression model, for retrieving the cloud property differences between the two years. \cite{kato14} defined the spectral difference as the anomaly from the long-term average spectral radiance, for retrieving atmospheric and cloud property anomalies from spatially and temporally averaged spectral radiance. 
In this study, we use the spectral radiance from the control run in \cite{kato11} as the functional response, and treat the spectral radiance from the $15$ perturbed runs with different cloud and atmospheric properties as the functional covariates, to investigate the piecewise linearity of their effects over space and the $24$ months, in order to study local features and grouped patterns. 
Subsequently, we regularize the sampling points with observed functional data that comprise the $16$ spectral radiances over the $2711$ wavenumbers onto a $2$-D regular lattice system, with the 18 $10^{\circ}$ latitude zones and the $24$ months from January 2003 to December 2004 as the coordinates.
For the regularized spatio-functional data, we seek for meaningful partitions of the $18\times 24 = 432$ spatial locations according to regional effects of the spectra radiances from the $15$ perturbed runs on the radiance from control run. The clusters under the partition are expected to be spatially contiguous 
for 
direct interpretation as subregions,
and more importantly for capturing the local small-scale variability with spatial dependence. 
Moreover, we target on not only identifying the heterogeneity of the effects among latitude zones and months, but also quantifying the significance and strength of such effects that can vary across the spectral regions of the wavenumbers. 
Finally, we 
intend to provide the uncertainty associated with such partitions, in particular, the variability among different partitions and the boundary ambiguity in the label assignment. 
We propose a 
model-based approach
for answering these questions in a unified faction.

\subsection{Related Work}
\label{subsec:liter}
The spectra radiance data in this analysis 
exhibit an intrinsic high-dimensionality. 
A well-known issue in partitioning large and high-dimensional spatial data sets
is the presence of nonstationarity or local stationarity in that the spatial patterns and behaviors of the underlying processes can vary across the study domain. 
Taking such local dependency into account can potentially improve both model fit and interpretability. 
One popular Bayesian clustering method is using Dirichlet Process prior (DPP) for horizontally partitioning the observations to identify common features in vertical direction. 
For example,  
\cite{rm06} combined horizontal partitions using DPP and vertical partitions using Bayesian wavelets for curve clustering with application to spatial data; 
\cite{dass} conceptually extended the feature vector for unsupervised learning to feature function, and elicited a functional DPP for simultaneously detecting change points for multiple curves over a spatial domain; among others.  
Nevertheless, these DPP-based approaches mentioned above do not guarantee that the resulting clusters are spatially contiguous.  
Whereas for spatial data,  the interpretation and capturing local stationarity can be greatly improved by seeking for spatially contiguous clusters, with natural interpretation as subregions.  
One 
convenient tool of introducing such spatially connected clusters is the treed Gaussian processes \citep{gl08} under Bayesian framework to stochastically search for treed partitions \citep{den98} which typically produce rectangle-shaped clusters. 
\cite{konomi2} adopted treed partition for multivariate Gaussian precesses which can potentially be applied to curve clustering. 
However, the number of parameters can rapidly grow when more clusters are introduced, and the penalization and dimension reduction in parameter space are not fully explored. 
Another convenient way of defining a spatial partition is using Voronoi tessellations (VT) by introducing a set of centers (nuclei) and certain rule for label assignment. 
\cite{okabe00} provided a comprehensive review of VT-type partitions with application in modeling spatial process.  
\cite{kr00} introduced a Bayesian clustering prior model based on VT for disease mapping to identify risk by regions. 
\cite{kmh05} considered piecewise stationary Gaussian processes based on VT-based partition to address nonstationarity issues. 
\cite{zlm14} developed a VT-based Bayesian hierarchical model for identifying spatial extreme in cancer disease. 
\cite{castro15} adopted VT-based Bayesian clustering model to estimate the neighborhood matrix for modeling areal count data. 
\cite{feng16} compared the performance of the VT-based spatial clustering with traditional approaches such as Poisson-CAR model for spatial count data, and demonstrated the merit of VT-clustering using several simulation studies and real data examples.  
VT-type partition is also useful in non-Bayesian, non-parametric procedure as a convenient tool for defining subregions to extract local representatives for post-hoc clustering \citep[see, for example,][]{sec13}. 
A brief comparison between treed partition and VT-based partition can be found in \cite{gl08} and \cite{zlm14}.  In general, 
treed partition seeks for axis-aligned splitting planes. 
VT-type partition puts less constraints on the shape of clusters and is suitable for identifying irregularly shaped patterns such as hot spots in disease. 
Nevertheless, VT-based approach still puts considerable constraints, such as convexity assumption, on the cluster shapes, which limit its use for clustering more complex objects.

The limitations of imposing strong constraints on cluster shapes for both treed and VT partition become even more severe for high-dimensional spatio-functional data. 
Suppose we seek for a partition of $N$ spatial locations, each with $T$ measurements. When $T$ is quite large comparing to $N$, 
without dimension reduction, switching a label for a single spatial location involves simultaneously testing $T$ hypothesis. Consequently, the constraints on cluster shapes result in a very low chance of accepting the change in 
label assignment
for even a few spatial locations.
This causes the well-known local-trap problem \citep{den98} in that the stochastic model search fails to accept any new clustering configuration, which hinders the thorough exploration for probable clustering configurations. 
Therefore, an extended VT-type partition with less constraints on cluster shapes is desirable for handling more complex structure in high-dimensional spatial data.

In addition, the cluster-wise mean can be explained by a set of covariates, and the covariates may have varying effects across the space. 
\cite{sanchez15} utilized covariates information for VT-based spatial clustering to identify subregions that share similarity in covariate effects. 
It is more challenging to investigate functional covariate effects that are potentially varying across spatial clusters for high-dimensional spatial data, since the introduction of new partitions can lead to a rapid growth in the number of parameters. The dimension reduction in the functional representation is desirable. One common technique is the Bayesian wavelets \cite[e.g.,][]{cg00, cg03, mc06} that partition the $T$-observations into multi-resolution, and the important signals are generally centering at the first several levels of resolution. 
Bayesian wavelets have been employed for 
identifying the cluster means of the functional responses via the wavelet representation, which introduces sparsity and decorrelation property 
\cite[e.g.,][]{rm06}. 
Nevertheless, the combination of Bayesian wavelets for functional covariates and spatially contiguous VT-type partition for spatial data has not been fully studied. 

The main contribution of this work is the development of a 
spatio-functional clustering (SFC) model with varying functional covariate effects across random subregions, and a generalized Voronoi tessellation (GVT) procedure for partitioning spatial data with less constraints on the cluster shapes. 
We view clustering as a horizontal partitioning operator on the data with $N$ locations, which results in spatially contiguous regions for interpretation and addressing nonstationarity issue for large spatial data; 
We also view the selection of important wavelet coefficients as a vertical, binary partitioning operator of the data with $T$ replicates to handle the high-dimensionality. 
Dimension reduction is achieved along both directions by introducing a two-component partitioning operator. The vertical partition is nested in the horizontal partition to identify spatial clusters that share identical or similar underlying characteristics.

The remainder of the article is organized as follows.  
In Section \ref{section:model}, we present a spatio-functional mixed-effects model with random groups. 
In Section \ref{section:cluster}, we propose a generalized Voronoi tessellation for partitioning spatial data with less constraints on cluster shapes. 
We briefly introduce the model implementation and other options in Section \ref{section:impl}. 
In Section \ref{section:real}, we provide the real data application on clustering the infrared spectral measurements in identifying regional and dynamic characteristics in atmospheric composition change, followed by some further discussions in Section \ref{section:conclusion}.

\section{A Spatio-Functional Clustering Model}
\label{section:model}

In this section, we present a spatio-functional model with clustering. 
Suppose for each location $s\in \mathcal{S}=\{1,2,\ldots,N\}$ in the spatial domain, we have $T$-by-$1$ response $\mY_s = (\mY_s(1), \ldots, \mY_s(T))'$ and a set of $T$-by-$1$ covariates $\mX_{si} = (\mX_{si}(1), \ldots, \mX_{si}(T))'$ for $i=1,2,\ldots,p$. 
Let $\varpi$ introduce a spatial partition of the domain $\mathcal{S}$ into $d$ contiguous clusters, i.e., $\mathcal{S}=\cup_{r=1}^d \cl_r$ with each cluster $\cl_r$ 
comprising
$n_r$ locations in $\mathcal{S}$, hence $\sum_{r=1}^d n_r = N$. 
Consider the following model for $s\in \cl_r, r=1,2,\ldots,d$, 
\begin{eqnarray}
\label{modelf}
\mY_s &=& \sum_{i=1}^p\quad\,\,\,
\underbrace{\mX_{si} \circ \mbeta_{ri}}_{\mathclap{\text{Hadamard product}}} + 
\mU_s +  \mepsilon_s
\end{eqnarray}
where $\mbeta_{ri} = (\mbeta_{ri}(1),\ldots, \mbeta_{ri}(T))'$ is the $T$-by-$1$ fixed-effect function for the $i$th covariate in the $r$th cluster, and $\mX_{si} \circ \mbeta_{ri}$ denotes its Hadamard (element-wise) product with $\mX_{si}$.
Note the Hadamard product can be alternatively written as $\tilde{X}_{si} \mbeta_{ri}$ if we let $\tilde{X}_{si}$ be the block diagonal matrix with each block $\mX_{si}(t)$; 
However, $\tilde{X}_{si}$ requires $\mathcal{O}(T^2)$ memory storage rather than $\mathcal{O}(T)$, and hence is used purely for our notational convenience. 
Additionally, note that $\mX_{si}$ includes an intercept for $i=1$, which represents the cluster mean level of the response $\mY_s$. 
The cluster-wise fixed-effect function captures the large-scale variation that can be explained by the functional covariates in addition to the cluster means. 
The location-specific random-effect function $\mU_s=(\mU_s(t_1),\ldots,\mU_s(t_T))'$ further captures the small-scale variation due to individual location in addition to the cluster mean. 
It can also measure the local (within-cluster) and global (between-cluster) spatial dependence. 
$\epsilon_s$ is the nugget function that captures the small-scale variation due to the noise.

While the regional effects $\mbeta_{ri}$'s are considered to capture the cluster-varying signals and local predictive relationship between $\mX_{ri}$'s and $\mY$, it is immediately seen from (\ref{modelf}) that the model complexity arises as a severe issue when the functional mixed-effects model incorporates the clustering structure. 
For instance, an extra cluster can result in at least $pT$ additional fixed-effect parameters. 
To avoid redundancy of parameters, we consider using the Bayesian wavelet smoothing technique for dimension reduction. 
We also utilize the well-known decorrelation property that yields simple covariance structure in the wavelet domain. 
More specifically, assuming $T = 2^J$, we let $\beta_{ri} = W\mbeta_{ri}$, where $W$ is the $T$-by-$T$ discrete wavelet transformation (DWT) matrix that induces an one-to-one map of the original data domain indexed by $t\in\{1,\ldots,T\}$ into wavelet domain that has multi-resolution structure with double-index $(j,k)$: $j=0,\ldots,J$ indexes the resolution level, and $k=1,\ldots,2^{j-1}$ indexes the location at each level $j>1$. 
Note that $j=0$ (hence $k=0$) corresponds to the scaling function that preserves considerable information in the data domain.  
More explicitly, let $W=((w_{lm}))_{1\leq l,m\leq T}$ with its $m$th $T$-by-$1$ column matrix denoted as $w_{\cdot m}$, and $l$th $1$-by-$T$ row matrix as $w_{l \cdot}$. 
Note $W$ is an orthogonal matrix with the inverse Wavelet transformation matrix $W^{-1}=W'$. 
Each element $\mbeta_{ri}(t) = w_{\cdot t}'\beta_{ri}$, i.e., 
\begin{eqnarray*}
\label{waveletdecomp}
\mbeta_{ri}(t) &=&\xi_{00}(t)  \beta_{ri}(00)+ \sum_{j=1}^{J}\sum_{k=1}^{2^{j-1}}\psi_{jk}(t)\beta_{ri}(jk)\\
\end{eqnarray*}
where $w_{\cdot t}$ consists of the wavelet basis functions $\xi_{00}(t)$ and $\psi_{jk}(t)$'s at level $j$ and location $k$ for $t=1,\cdots,T$. 
Note $w_{\cdot t}' \neq w_{t\cdot}$ since $W$ is not symmetric, otherwise $W=W'=I_T$. 
The dimensions at both sides under this transformation match as $1+\sum_{j=1}^{J}\sum_{k=1}^{2^{j-1}} = 2^J = T$. 
The dimension reduction can be achieved by assuming a large portion of the wavelet coefficients $\beta_{ri}(jk)$'s have trivial or rather weak signals towards $Y_s$'s, and hence can be treated as exactly zero, in particular for larger $j$. 
Subsequently, we consider the model (\ref{modelf}) in the wavelet domain by multiplying $W$ at both sides of the equation:
\begin{eqnarray}
\label{model0}
Y_s &=& \sum_{i=1}^p
X_{si}  \beta_{ri} + u_s +  \epsilon_s
\end{eqnarray}
where $Y_s=W\mY_s$, $\beta_{ri} = W\mbeta_{ri}$, $u_s=W\mU_s$, $\epsilon_s=W\mepsilon_s$, and $X_{si}=W\tilde{X}_{si}W'$. 
Note albeit $\tilde{X}_{si}$ is diagonal, $X_{si}$ can be dense, causing aforementioned memory issues. 
By decorrelation property of the wavelet transformation, we assume 
$\epsilon_s\indsim \mathcal{N}(0, \, \sigma_r^2M_r)$ for the $r$th cluster, where $M_r$ is a diagonal matrix with $T$ diagonal entries $(m_{rjk})$ to measure the heterogeneity of the noise level at different wavelet basis locations $(j,k)$. 
We fix $m_{r00}\equiv 1$ such that $\sigma^2_r$ represents the noise level for the scaling function $(j,k)=(0,0)$, and $m_{rjk}$ is the relative variability  at individual level $j$ and location $k$ for data to the base variability $\sigma^2_r$ for the $r$th cluster. 
Note that as long as $M_r\neq I_T$, the residuals $\mepsilon$ in the data domain are correlated with covariance $\sigma_r^2WM_rW'$. 
It is also convenient to write (\ref{model0}) as
\begin{eqnarray}
\label{model1}
Y_s &=& \sum_{i=1}^p \sum_{j,k}
X_{si}(jk)  \beta_{ri}(jk) + u_s +  \epsilon_s 
\end{eqnarray}
where $X_{si}(jk)=(X_{si1}(jk),\ldots,X_{siT}(jk))'$ is the $\tau$th column of $X_{si}$ 
that corresponds to $(j,k)$ for $\tau\in\{1,\ldots,T\}$; 
Its $t$th element, $X_{sit}(jk) = \sum_{l=1}^Tw_{t l}w_{\tau l}\mX_{si}(l)$, 
has the contribution $\beta_{ri}(jk)$ towards $Y_s(t)$. 
We routinely adopt the spike-and-slab prior for each $\beta_{ri}(jk)$ with structured parameters:
\begin{equation}
\label{priorbeta}
\beta_{ri}(jk)\given \varpi, \ugamma \indsim  \mathcal{N}(0, \,  \sigma_r^2 \lambda_{rij} \gamma_{ri}(jk)), \,\,\,
\gamma_{ri}(jk) \indsim \mbox{Bernoulli}\, (\pi_{rij})
\end{equation}
where $\lambda_{rij}$ is the scaling parameter, or signal-to-noise ratio of $\beta_{ri}(jk)$ versus the noise level $\sigma_r^2$ at $j=0$; $\gamma_{ri}(jk)$ is an indicator that if $\beta_{ri}(jk)$ can be treated as exactly zero: when $\gamma_{ri}(jk)=0$, the prior $\beta_{ri}(jk)\given\varpi, \gamma_{ri}(jk)$ is a point mass at $0$. 
The hierarchy of $\gamma_{ri}(jk)$ depends on the prior shrinkage probability $\pi_{rij}$, which allows Bayesian learning of the shrinkage effects for the extent of dimension reduction. 
The horizontal partition operator $\varpi$ and vertical partition operator $\ugamma$ that consists of all $\gamma_{ri}(jk)$'s, jointly determine the 
number of nonzero elements, or $\ell_0$-norm of 
$\ubeta$ coefficients. 
We let $q_{ri}=\sum_{jk}\gamma_{ri}(jk)$ be the $\ell_0$-norm of $\beta_{ri}$. 
It is possible to consider Ising (auto-logistic) prior \citep[see, e.g.,][]{sf07} for $\pi_{rij}$'s in (\ref{priorbeta}), if we further introduce a coarse-level neighborhood structure for the spatial clusters. However, for a relatively small number of clusters, the spatial dependence between cluster-wise parameters can be weak.
 We alternatively incorporate the Ising-type spatial averaging for the wavelets shrinkage parameters in the proposal densities when creating new clusters.

As remarked by \cite{mc06}, allowing the scaling parameter $\ulambda$ to also depend on the location $k$ introduces extra flexibility and improves the potentiality of capturing signals at finer scales. 
In the proposed model with clustering, the number of the scaling parameters can, however, proliferate with the flexible assumption. 
We instead allow the scales $\lambda_{rij}$'s and shrinkage levels $\pi_{rij}$ to vary across clusters for capturing more details via borrowing regional information. 
This assumption adds more flexibility to that in 
\cite{rm06}, where the authors assumed homogeneous scaling and shrinkage levels of wavelet functions over clusters.  
\cite{gl08} also considered cluster-specific hyperparameters for the treed Gaussian Processes.  

For the spatial random effect function $u_s$, a common assumption is that the correlation structures within and between $u_s$'s are separable \citep[e.g.,][]{mc06}, such that the wavelet transformation only applies to the within-correlation structure, and the between-curve correlation structure is preserved in the wavelet domain. 
For multivariate spatial data such as temporal or functional response with treed partitions of the study domain, \cite{konomi} assumed separable covariance functions for measuring within-unit interactions of the multivariate response and between-unit spatial dependence. 
Despite bringing considerable advantage in computation, the assumption of separable covariance function has known issues \citep[e.g.,][]{stein05}. 
Moreover, for wavelet coefficients that contain distinct strengths of signals, it is generally not evident that the spatial dependence of the corresponding scale and shrinkage parameters is common. 
We therefore consider a non-separable dependence structure. 
For notational convenience, we denote $\cl_r=\{1,2,\ldots,n_r\}$ 
i.e., $\cl_r$ consists of the first $n_r$ locations.  
Let $u_{rjk}=(u_1(jk), \ldots, u_{n_r}(jk))'$ inherit the decorrelation assumption in wavelet domain, we consider:
\begin{equation}
\label{prioru}
u_{rjk}\given \varpi \indsim  \norm\left(\zero_{n_r}, \,  \sigma_r^2 H_{rjk}\right), \,\,\,
H_{rjk} = h_{rjk}  (F_r-\phi_{rjk} Q_r)^{-1} 
\end{equation}
where we assume the conditional autoregressive (CAR) structure \citep[see, e.g.,][]{bym91} for the spatial dependence among $\tilde{u}_s(jk)$'s for $s\in \cl_r$: $h_{rjk}$ is the scaling parameter, or detectability of the random effect $u_{rjk}$ comparing to the noise level $\sigma^2_r$ at $j=0$; 
$Q_r$ is the neighborhood matrix for cluster $\cl_r$, with diagonal entries equal $0$. 
$F_r$ is a diagonal matrix with each entry being the sum of neighbors. 
$\phi_{rjk}$ is the parameter that measures the spatial dependence for wavelet location at $(j,k)$ in $\cl_r$. To guarantee the positive-definitness of $H_{rjk}$, $\phi_{rjk}$ has finite support $(1/\rho_{rn_r}, 1/\rho_{r1})$ where $\rho_{r1}$ and $\rho_{rn_r}$ are the largest and smallest eigenvalue of $F_r^{-1/2}Q_rF_r^{-1/2}$, respectively.  
Note (\ref{prioru}) specifies a localized random effect within each cluster, or a random effect nested in the cluster-specific random effect $\beta_{r1}$. It is possible to assume a global spatial random effect that further introduces the dependence between curves in different clusters. 
However, this depends on the computational efficiency and memory storage, since the computational burden explodes as $\mathcal{O}(N^3)$. 
The localized random effect hence can greatly improve the computational efficiency in particular when $N$ is large.   
Consequently, the support of each spatial parameters $\phi_{rjk}$ depends on the partitions introduced by $\varpi$ since $F_r$ and $Q_r$ may vary, which adapts the prior knowledge under different partitions. This is another different aspect from the existing approaches on spatial partitioning models which typically assume a global prior for the spatial parameters \citep[e.g.,][]{kmh05, gl08}. 

To summarize, given a random horizontal partition operator $\varpi$ that groups data into cluster $\cl_1, \ldots, \cl_d$, and a random vertical binary partition $\ugamma$ that suggests $q_r$ nontrivial signals for cluster $\cl_r$ with $\sum_{ijk}\gamma_{rijk}=q_r$,  
we propose a spatio-functional clustering (SFC) model 
for $n_r$ functional response
$Y_{rjk}=(Y_{1jk}',\ldots, Y_{n_rjk}')'$ with corresponding $n_r$-by-$q_r$ design matrix $X_{rjk}$: 
\begin{flalign}
\label{prior_v} &\varpi \sim \pi(\varpi), \\
\notag &\gamma_{ri}(jk) \indsim  \mbox{Bernoulli}\, (\pi_{rij}), 
\quad \pi_{rij}\iidsim \mbox{Beta}\left(a_{\pi, ij}, \, b_{\pi, ij}\right), \\
\label{lik_r}  &Y_{rjk}  \indsim \norm\left(X_{rjk}\beta_r + u_{rjk}, \,\, \sigma^2_r M_{rjk} \right), \quad  M_{rjk}=m_{rjk}I_{n_r},\\
\notag &m_{rjk}\indsim\igamma(a_{m,jk}, b_{m,jk}), \quad m_{r00}\equiv 1,\\
\label{prior_b} & \beta_r \indsim \norm\left(\zero_{q_r},\,\, \sigma^2_r\Lambda_r\right), 
\quad \Lambda_r = \mbox{diag}_{ijk\in\{(i,j,k): \gamma_{ri}(jk)=1\}} \lambda_{rij},\\
\notag &\lambda_{rij} \indsim \igamma\left(a_{\lambda,ij}, b_{\lambda,ij}\right), \\
\label{prior_u} &u_{rjk}\indsim \norm\left(\zero_{n_r},\,\, \sigma^2_rH_{rjk}\right), \quad 
H_{rjk} = h_{rjk}  (F_r-\phi_{rjk} Q_r)^{-1},\\
\notag &h_{rjk}\indsim\igamma\left(a_{h,jk}, \, b_{h,jk}\right), \\
\notag &\phi_{rjk} \indsim \unif(1/\rho_{rn_r}, 1/\rho_{r1}), \\
\label{prior_e}& \sigma_r^2 \iidsim \igamma\left(a_{\sigma}, \, b_{\sigma}\right).
\end{flalign}
This prior specifications in the full Bayesian hierarchical model above introduce a set of hyper-parameters for the wavelets: the signal-to-noise ratio $\lambda_{rij}$, and the probability of shrinkage $\pi_{rij}$, which can be routinely estimated using an empirical Bayes procedure as described 
by \cite{cg00} and \cite{mc06}. 
However, the empirical Bayesian estimation procedure becomes intractable when the random clustering structure is involved. Furthermore, the clustering results can be sensitive to the choice of these hyperparameters. 
We thereby pose a further hierarchy to update these parameters, and draw posterior samples for cluster-wise inference. 
For all Inverse-Gamma prior densities, we choose the hyper-parameters to yield rather dispersed prior density, such as shape $a = 2$ and $b=0.01$. 
Note that for the variance term $\sigma_r^2$, the choice of $a_{\sigma}=0$ and $b_{\sigma}=0$ corresponds to the noninformative, scale-invariant Jefferys prior which may affect the expression of the marginal model likelihood. 
For Beta prior one can choose $a=b=1$ to yield the uniform prior on $(0,1)$ for $\pi_{rij}$, which does not violate the posterior propriety since $\pi_{rij}$ has bounded support while the outcomes of $\gamma_{ri}(jk)$'s form a finite set. 
The full and marginal model likelihood is discussed in Appendix \ref{section:appenA}.

\section{Generalized Voronoi Tessellation for Spatial Clustering}
\label{section:cluster}
In this section, we discuss the choice of $\pi(\varpi)$ in (\ref{prior_v}), and couple the proposed SFC model with a generalized spatial partitioning model. 
Let $\varpi$ be a clustering configuration introduced by a Voronoi tessellation \citep{okabe00, kr00, kmh05, zlm14}. It is determined by two components, $\varpi=(d, \, L)$, with $d$ the number of clusters, and $L=(\ell_1,\ell_2,\cdots,\ell_N)$ the label vector that indicates the membership for each location, $\ell_s\in\{1,2,\cdots,d\}$. Note we have suppressed the implicit dependence of $L$ on $d$ for notation simplicity. 
For spatial clustering, we can reduce to $\varpi=(d, \, \cen_d)$ where $\cen_d = (g_1,\cdots, g_d)$ are cluster centers and $g_r\in \mathcal{S}$. The label assignment $L$ is determined by the minimal distance criterion, i.e., locations with the minimal distance from center $g_r$ form $\cl_r$. 
Hence we have each cluster $\cl_r=\{s\in\mathcal{S}: \ell_s=r\} = \{s\in\mathcal{S}: D(s, g_r) \leq D(s, g_k), \forall g_k\in \cen_d\}$. 
For point reference or irregularly spaced data, one natural choice for $D(\cdot, \cdot)$ is the Euclidean or the great circle distance between the spatial coordinates;
For areal or regular lattice data that can be regularized on a graph, for each location $s$, we define the order of a neighbor location $s'$ as the number of boundaries to cross from location $s$ to location $s'$, and use it as the distance $D(s,s')$. 
Note that the centers that correspond to the minimal distance may not be unique, hence $\cen_d$ does not uniquely induce a label assignment $L$. 
A remedy by \cite{kr00} is to treat $\cen_d$ as an ordered set, 
and 
when the minimal distance for location $s$ corresponds to more than $1$ centers including $g_r$, 
the procedure consistently assigns $s\in \cl_r$ 
if $g_r$ 
has the smallest index in $\cen_d$. 
The stochastic model search then involves swapping the center index and comparing the overall model likelihoods to check which assignment is more appropriate. 
However, the locations with minimal distance ties are still simultaneously assigned to either cluster regardless of the individual likelihood. 
This depicts the limitations of VT-based partition in handling more complex structures. 

A frequently adopted clustering prior model \citep[e.g.,][]{kr00} is to define $\pi(\varpi) = \pi(d)\,\pi(\cen_d\given d)$, where $\pi(d)$ is a prior model for the number of clusters $d$, and $\pi(\cen_d\given d)$ is a prior model for the cluster centers, $\cen_d$. 
One may specify a minimal size $n_0$ such that $n_r\geq n_0$ for all $\cl_r$'s, especially in the presence of covariates. 
Consequently $d$ is bounded above, say $d\leq N_0\leq N$. 
Consider the truncated geometric prior $\pi(d\given\alpha) =C(\alpha)(1-\alpha)^{d-1}$ for $d=1,\cdots,N_0$ with the normalizing constant $C(\alpha) = \alpha/(1-(1-\alpha)^{N_0})$ with tunning parameter $\pi(\alpha) \sim \unif(0,1)$. 
Note that when $\alpha$ is larger, $d$ is more likely to be smaller, while when it is fixed at a very small value, $d$ is almost uniformly distributed over the grids $\{1,2,\cdots,N_0\}$, such that each possible value of $d$ receives the same prior weight. 
Therefore, the hyperparameter $\alpha$ allows the flexibility to introduce $L_0$-type penalty on redundant parameters \citep[see][]{zlm14}. 
Conditioning on $d$, a uniform prior on centers is specified as $\pi(\cen_d\given d) = (N-d)!/ N!$ 
such that all possible $\cen_d$'s with $d$ centers receive equal prior weight. 
Comparing to the treed partition which typically aligns the boundary of clusters to the axes, 
the partition induced by Voronoi tessellation puts less constraints on the shape of clusters. 
However, it still imposes quite strong constraints such as convexity, which limits its use for more generally and irregularly shaped patterns of clusters, and for capturing the boundary ambiguity which is a well-known issue in image segmentation.  

Therefore, we extend a clustering model with a generalized Voronoi tessellation (GVT) by introducing the boundary concept, and assuming a prior distribution on the labels of a boundary set. 
Let $\varpi = \{d, \, \cen_d, \, L\}$ where $d$ is the number of clusters, $\cen_d = \{g_1, \ldots, g_d\}$ is a set of cluster centers which determines a partition of $\{1,\ldots,N\} = \cup_{1\leq r\leq d} \cl_r$ for the spatial domain, and $L = \{\ell_s, s=1,\ldots,N\}$ is the label assignment where $\ell_s \in \{1,2,\ldots,d\}$. 
We define a boundary set $\bds\subset \mathcal{S}$ induced by $\cen_d$ in the following two steps
for each location $s=1,\ldots,N$: 
\begin{itemize}
\item[1.] Let $K_s = \{r: D(s, g_r) = \min_{1\leq m\leq d}D(s, g_m)\}$. If its cardinality $\mathbf{card}(K_s) = |K_s| >1$ 
i.e., location $s$ has the minimal distance from at least $2$ cluster centers, then location $s$ is a boundary point. Additionally, it is referred as a tied point; 
\item[2.] Next, for some integer $K$, suppose all the neighbors of location $s$ with orders $\leq K$, except tied points, form a set $\mathring{\mathcal{N}}_K(s)$. 
Let $K_s$ be the set of unique labels that the locations in $\mathring{\mathcal{N}}_K(s)$ receive. 
If $|K_s| >1$, then location $s$ is a boundary point. 
\end{itemize}
In either case, $K_s$ is called the {\em choice set} of the boundary point $s$.
Consequently, the boundary set $\bds$ is the collection of all such boundary points. 
In general, one may pick smaller $K$ values to specify rather thin buffer regions near the boundaries to avoid fuzziness that hampers the spatial contiguity and hence the interpretability of clusters. 
Note for a non-boundary point, or an interior point $s \in \mathcal{S}\setminus  \bds$ or $\bds^c$, its choice set only contains the cluster label assigned to location $s$, i.e., $K_s = \{r: D(s, g_r) = \min_{1\leq m\leq d}D(s, g_m)\}$ since $|K_s|=1$.  

Next, we specify the clustering model prior as $\pi(\varpi) = \pi(d)\,\pi(\cen_d\given d)\,\pi(L\given \cen_d)$ with independent label assignments $\pi(L\given \cen_d)=\prod_{s=1}^N\pi(\ell_s = r \given \cen_d)$. 
For instance, under the uniform prior weights, we have $\pi(\ell_s = r \given \cen_d) = 1/|K_s|$ for label $r\in K_s$. Other choices include using the proportion of members in the neighbor set of $s$ that receive label $r$ as the prior weight $\pi(\ell_s = r \given \cen_d)$. However, such majority rule can pose a strong assumption against the true structure suggested by the data. We hence stick to the simple, uniform prior weights to avoid additional computational effort. 

Spatial clustering techniques based on Voronoi tessellation in previous studies \citep[e.g.,][]{kr00, kmh05, sec13, zlm14} choose the minimal-distance criterion and minimal-index-assignment rule by treating $\cen_d$ as an ordered set, to restrict $|K_s|=1$ for every location $s$ for making binary assignment. 
In our case, $|K_s|=1$ holds only for the interior points $s\in \bds^c$, and we assume random assignments for the boundary points. 
The implementation hence allows posterior boundary label correction for $L_{\bds}=\{\ell_s, s\in \bds\}$ given local cluster parameters, and results in more generally shaped clusters to incorporate the boundary ambiguity.  
The extended clustering model prior can be more concisely presented as 
$\pi(\varpi) = \pi(d)\,\pi(\cen_d\given d)\,\pi(L_{\bds}\given \cen_d)$ 
and we prefer $\pi(\cen_d\given d) \propto \prod_{s\in \bds}|K_s|$ as described in Appendix \ref{section:appenB}.

\begin{figure}[!ht]
\centering
\hspace{.7cm}\includegraphics[scale=.49]{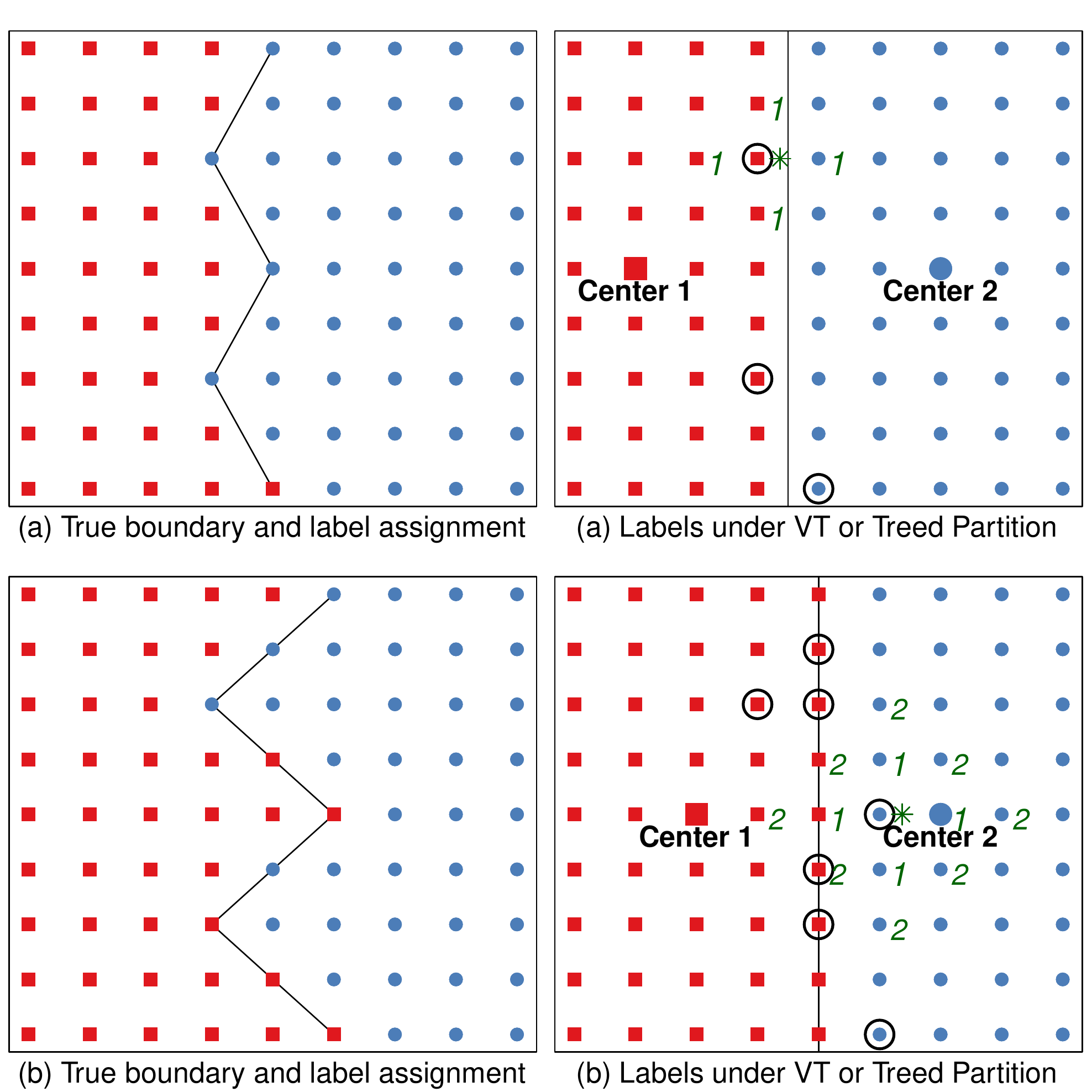}
\vspace{-.3cm}\caption{ \label{fig:E} 
Two examples of spatial partitions with boundary correction. 
}
\end{figure}

We illustrate the flexibility using boundary correction with two examples in Figure \ref{fig:E}. 
Let $\mathcal{I} = \{1,2,\ldots,9\}$ and we consider the $81$ locations on the regular lattice system $\mathcal{I}\times \mathcal{I}$ with two clusters.  
In Example (a), the top-left panel shows the true boundary and label assignment which potentially describe the dynamic boundary change due to the seasonal effect in the spectra radiance study. 
Next, the top-right panel shows one example partition using Voronoi tessellation based on either Euclidean distance or shortest path with 4-point neighborhood system. In this case, the regular VT-based partition \citep[e.g.,][]{kmh05} agrees with treed partition \citep[e.g.,][]{gl08} in producing a boundary that aligns to the axis, and the limitation is immediately seen for the misclassified locations near the boundary, as highlighted with circles. 
However, with boundary correction, these misclassified locations can still receive the correct label by comparing the likelihoods evaluated at both sides. 
Take the location with an asterisk for example, its order-$K$ neighbors receive different labels with $K=1$ hence it is a boundary point with the label determined by the likelihood. 
In example (b), the bottom-right panel visualizes a partition using Voronoi tessellation that introduces distance ties in that some locations have the same minimal distance to both cluster centers. 
In this case, both the regular VT and treed partition typically assign the boundary points simultaneously to either side based on some ordering of the centers. 
The boundary correction with $K=2$ allows point-wise label correction adjusted by the likelihood for the tied points. 
Note that one can assume random $K$ with certain prior distribution. 
A large $K$ not only results in additional computational effort in correcting larger boundary set, but also potentially hampers the spatial contiguity. Since the spectra radiance data set does not exhibit clear depth in boundary ``invasion'' and high complexity in shape, we fix $K=2$, while other similar choices do not severely affect the results from our sensitivity analysis. 
In addition, the boundary flexibility under the new partitioning operator substantially accelerates the chance of traversing in the model space for the stochastic search.

\section{Model Implementation}
\label{section:impl}
The model implementation involves the stochastic search of the horizontally partitioning operator $\varpi$ for clustering, with associated vertically partitioning operator $\ugamma$ for variable selection and the model parameters two layers: 
(1) higher-layer parameters $\theta_1 = \{ m_{rjk}, \lambda_{rij}, \pi_{rij}, h_{rjk}, \phi_{rjk} \}$ that measure the scale, dependence, and shrinkage for the wavelet components, and
(2) lower-layer parameters $\theta_2 = \{u_r, \beta_r, \sigma_r^2\}$ that can be routinely integrated out and the resulting marginal likelihood has a tractable form, as given in Appendix~\ref{section:appenA}. 
The estimation of the parameters follows a standard procedure under Bayesian framework using Markov chain Monte Carlo (MCMC) technique for posterior inference. 
Similar to many previous studies \citep[e.g.,][]{kr00, feng16, zlm14} that utilized the reversible jump MCMC algorithm, the stochastic search for partitions generally involves a split-merge step in transdimensional move for model exploration. In addition, we consider the model exploration in the same dimensional space, which involves the stochastic boundary search and  boundary correction under our new partitioning model. The details are given in Appendix~\ref{section:appenB}.    

It is worth pointing out that \cite{jasra07} considered a population-based reversible jump MCMC algorithm to alleviate the local-trap problem in multi-modal model space exploration. The idea is to allow the communications between multiple rj-MCMC runs by assigning a temperature ladder to pave the gap between the marginal likelihoods among chains.  The authors also utilized the delayed rejection \citep{gm01} for accelerating the acceptance rate.  
On the other hand, \cite{kmh05} employed the population-based Evolutionary Monte Carlo (EMC) \citep{lw01} for searching spatial partitions based on the regular Voronoi tessellation, as an alternative to the rj-MCMC algorithm. \cite{br10} also developed the EMC sampler for tackling high-dimensional model search issues in Bayesian variable selection, which is attractive to our application for vertical partitioning of wavelet coefficients. 

One limitation of the population-based MCMC algorithm is that the communication inner chains can undermine the pure independence which is ideal for parallel computing over MCMC runs. 
Since the full Gibbs step within each chain can be quite time-consuming in particular for updating the components associated with the fixed-effect functions, we instead adopt two-level parallel computing scheme. 
Firstly, we parallelize the potentially quite large number of MCMC runs. 
Secondly, we parallelize the subsequent Gibbs sweep of model parameters for each subregion after sampling the partitioning operator within each MCMC run.   
The two-level parallelism turns out to substantially facilitate the SFC implementation in this data application. 
Next, to alleviate the local-trap issue, we start with pilot runs with a large number $L=500$ MCMC chains for recording locally-trapped partitions, and use them as the ``anchor'' points in the model space for the subsequent full stochastic search that utilizes the delayed rejection algorithm \citep{gm01}. The performance is evaluated by checking if a large portion of distinct sampled partitions are presented in the posterior samples.

\section{Application to Spectra Radiance Data}
\label{section:real}
\begin{figure}[!ht]
\centering
\includegraphics[scale=.77]{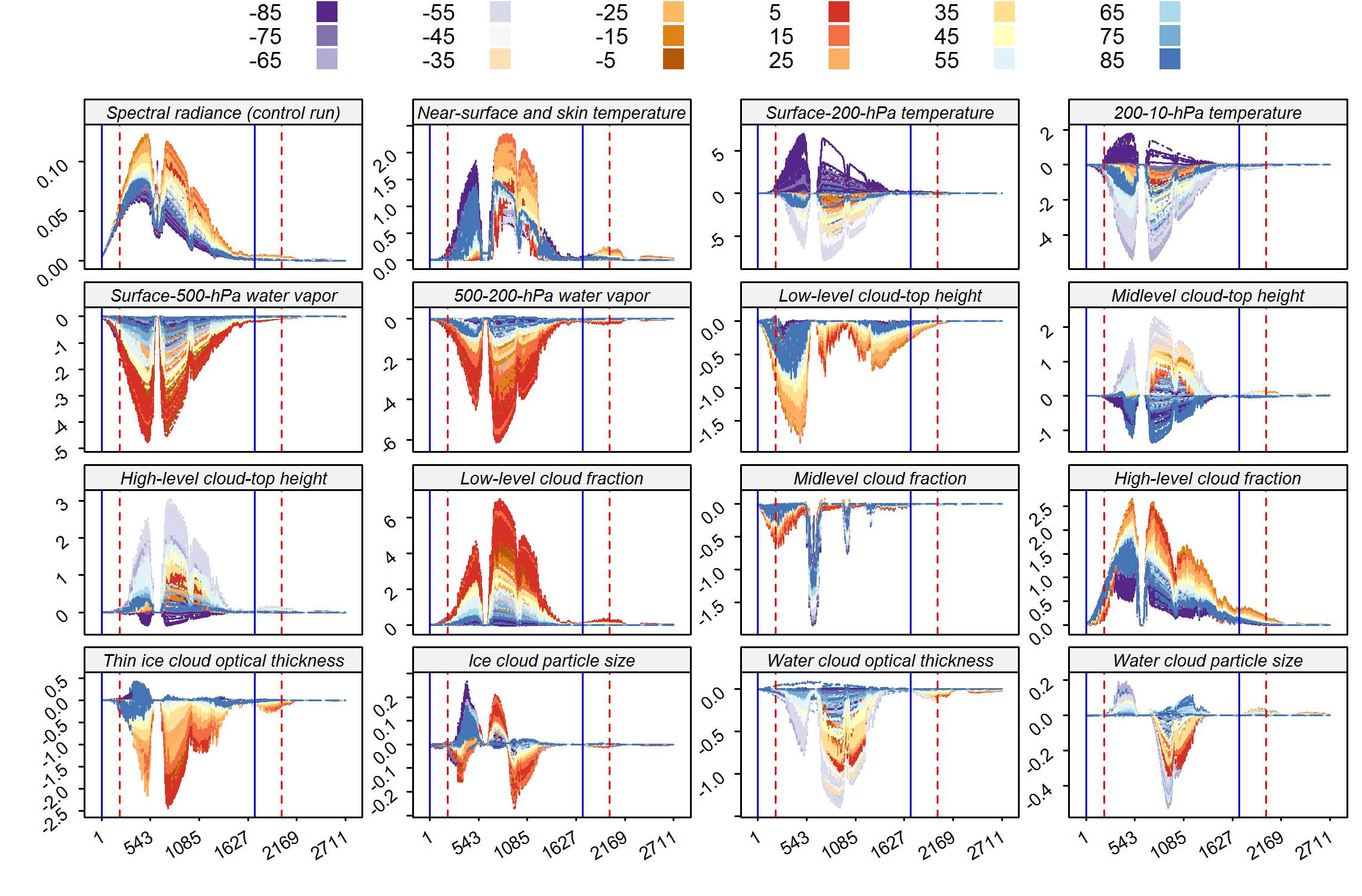}
\vspace{-.9cm}
\caption{ \label{fig:All} 
Spectral radiances over wavenumber (cm$^{-1}$). The 15 perturbed runs were multiplied by $10^4$ from the original data, to have similar order of magnitude as the control run.
}
\end{figure}

Top-of-the-atmosphere (TOA) longwave nadir-view spectra from $50$ to $2760$ cm$^{-1}$ are computed with a $1.0$ cm$^{-1}$ resolution for 18 $10^{\circ}$ latitude zones and $24$ months from January 2003 to December 2004. 
The raw data over the observed $2711$ wavenumbers for the $18\times 24 = 432$ curves, including one control run (top-left panel) and $15$ perturbed runs using indicated cloud and atmospheric properties, are shown in Figure \ref{fig:All}. 
The $432$ curves have meaningful colors indicating their $10^{\circ}$ latitude zones with the central latitude shown on the top of Figure \ref{fig:All}. 
The dark purple color indicates the 24 monthly radiances from the south pole latitude zone ($-90^{\circ}$ to $-80^{\circ}$), while the dark blue color indicates the radiances from the north pole latitude zone. 
On the other hand, the dark red and brown color indicate the radiances from latitude zones near the equator. 
The response radiances (control run) clearly exhibit spatially contiguous group patterns that motivate this study. The near-pole zones evidently have smaller means and variabilities for the radiance, while the near-equator zones have higher quantities. Nevertheless, the covariates information can be different for the two near-pole zones, and it is therefore important to study the zonal effects rather than simply clustering accordingly to the means.

We apply the proposed SFC model with the spatial partitioning method to 
the spectral radiance data, by treating the radiances from control run as the response, 
and the output from the $15$ perturbed runs as covariates. 
On the $2$-d lattice system regularized by month and latitude, 
We adopt the Manhattan distance, or equivalently the shortest path in this case, $D$ between a pair of locations $s_1$ and $s_2$, which are neighbors if $D(s_1, s_2) = 1$. 

The original number of $T=2711$ wavenumbers with the total $16$ runs can be relatively large and slow down the full implementation of component-wise updating the fixed-effects wavelet coefficients during the MCMC runs. 
Additionally, the primary goal is to identify the grouping structure.
Thus, we consider preprocessing for dimension reduction by systematically sampling $T=128$ wavenumbers from $1$ to $1700$ cm$^{-1}$ with resolution around $13$ cm$^{-1}$. 
Other pre-regularization techniques such as using P-splines or kernel functions can be also considered if the resulting curves are shape-preserving and not over-smoothing the raw curves.  
The interval is empirically determined by checking the signals for the spectra radiance, and is largely overlapping with the interval ($200$ to $2000$ cm$^{-1}$) in \cite{kato11} for calculating the relative difference. 
The interval between $1700$ and $2000$ cm$^{-1}$ is dropped in this analysis since the corresponding radiance for the control run is quite flat, close to 0. 
Furthermore, all the functional covariates are multiplied by $10^4$ since the original scales are quite small comparing to the response radiance from the control run. 
For each of $N=24\times 18 = 432$ locations, we have $T=128$ wavenumbers for each of the $16$ variables. 
As a consequence, we have a total of $432\times 128\times 16 = 884,736$ data points.

The data for this analysis are shown in Figure \ref{fig:All}. The two red vertical lines specify the window including the fraction of data used by \cite{kato11}. 
The two blue vertical lines specify the window including the fraction of data used in this analysis. We further perform a systematic subsampling to reduce the number of wavenumbers to be $128$, with a coarser level of resolution of $13$ cm$^{-1}$ to accelerate the horizontal partition search.

After pilot runs for initial partitions and ``anchor'' points for model exploration, 
we fit the Bayesian model with $5$ MCMC runs with distinct initial values and a total of $10,000$ iterations per chain. 
At each iteration, the clustering configuration $\varpi$ is updated, and then the parameters for each subregion are updated via 
{\tt MATLAB} R2014a parallel computing toolbox with 16 cores per chain. 
The computational benefit with parallel computing depends on the number of subregions $d$, and is generally significantly achieved comparing to the single-thread runs. 
On average, $1,000$ iterations take approximately an hour to complete. 
The most computational expensive part is updating the $d\times p\times T$ vector of $\beta_r$'s which is conducted sequentially and component-wise due to the wavelet smoothing prior for dimension reduction.

\begin{figure}[htbp]
\centering
\includegraphics[scale=.7]{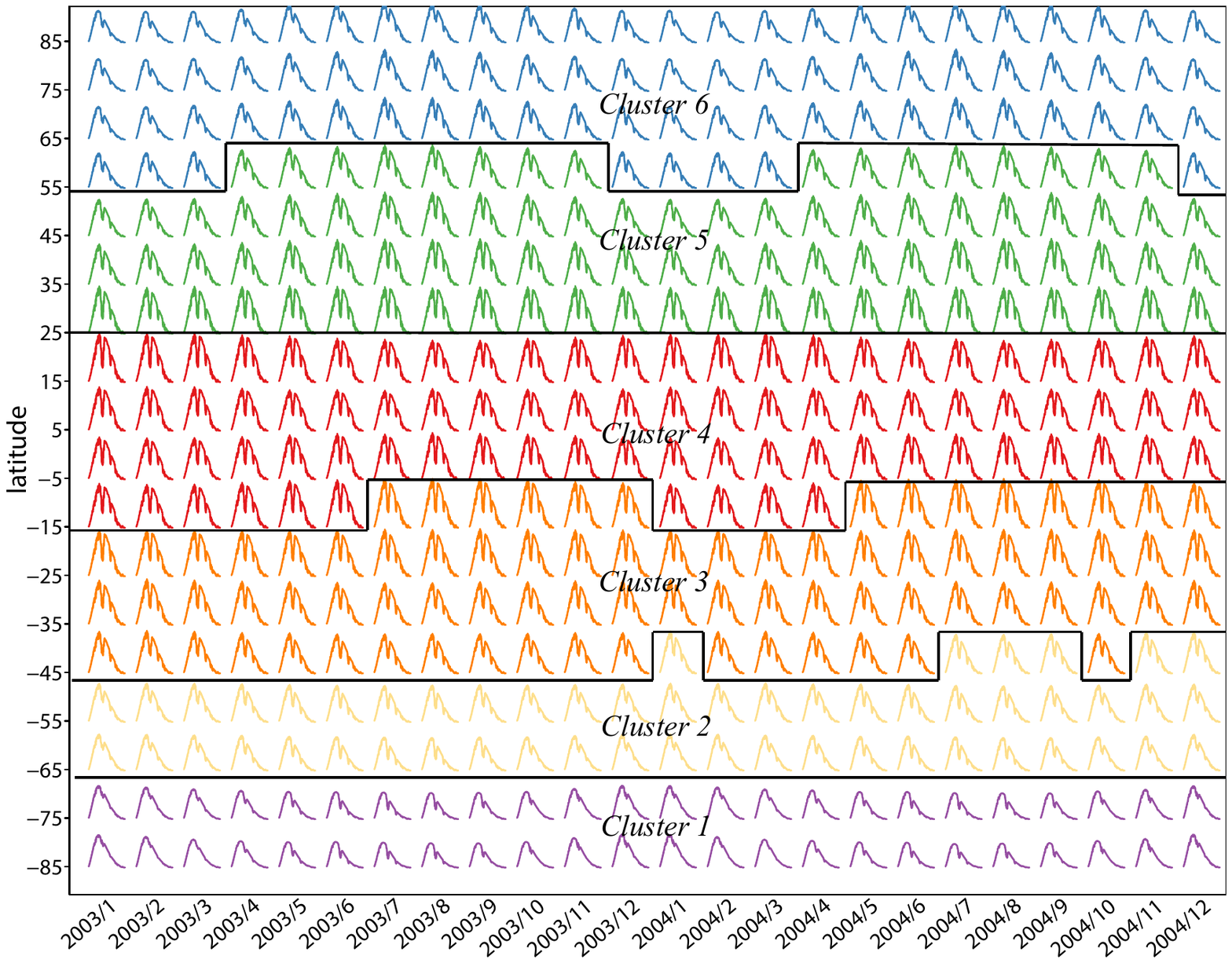}
\vspace{-5.6cm}
\caption{ \label{fig:L} 
The maximal a posteriori (MAP) estimate of the clustering configuration. 
}
%
%
%
\vspace{1cm}
\centering
\includegraphics[scale=.77]{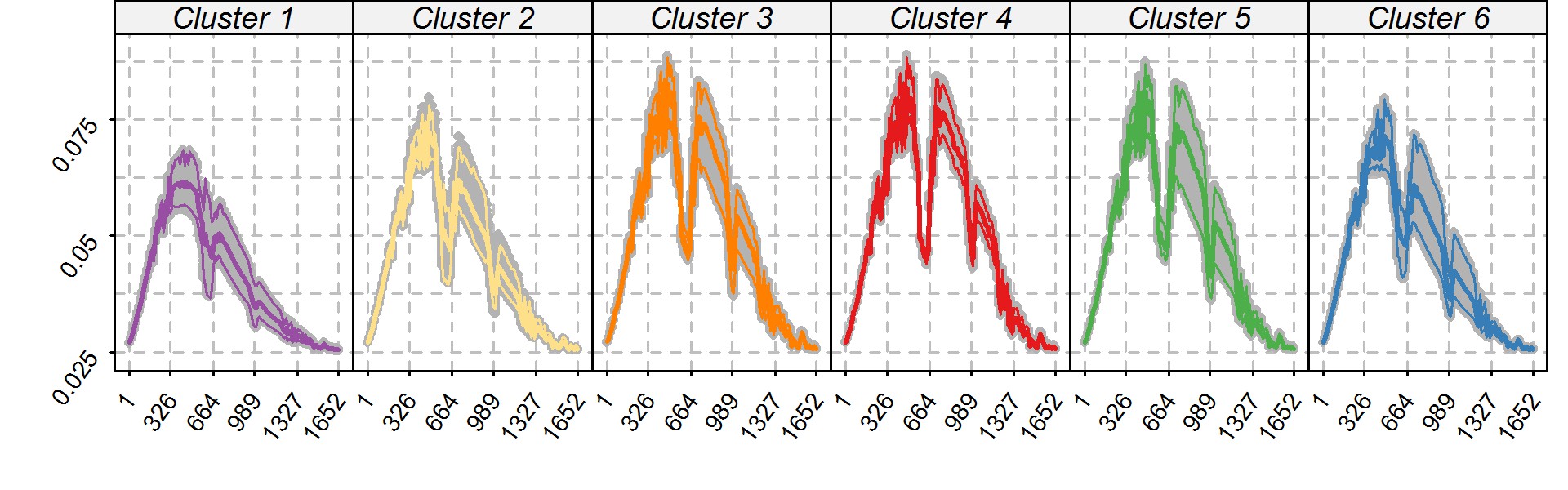}
\vspace{-.7cm}
\caption{ \label{fig:LY} 
Cluster-wise summary for $\hat{\mY}_s$ over wavenumber (cm$^{-1}$).
}
\end{figure}

The posterior modal estimate $\hat{\varpi}$ that partitions the response spectral radiance from control run, and the corresponding posterior estimates of the mean curve with $95\%$ predictive bands for each cluster, are shown in Figure \ref{fig:LY}. 
Since we regularized the observed spectral radiance into month$\times$lattice 2D lattice system, the map in Figure \ref{fig:LY} can be viewed as a dynamic clustering map for the $10^{\circ}$ latitude zones over the $24$ months in $2003$--$2004$. It is evident to see how the clustering map evolves over months.
By incorporating the information from $15$ atmospheric compositions as functional covariates, the trends and variabilities of the spectral radiances can be well captured. The posterior mean and the $95\%$ predictive band for cluster-wise mean curve are computed using $\hat{\mu}_{rjk}^{(b)} = 
\sum_{s\in \cl_r}\left(
 \sum_{i=1}^p\sum_{j,k}X_{si}(jk)\beta_{ri}(jk)^{(b)} + u_s^{(b)} 
\right)/\hat{n}_r$
for all posterior samples $b$ that has $\varpi^{(b)} = \hat{\varpi}$, and $\hat{n}_r$ is the number of locations in cluster $\cl_r$ under $\hat{\varpi}$. 
The curves are transformed back to the original data domain for visualization and interpretation.

\begin{figure}[!ht]
\centering
\includegraphics[scale=.77]{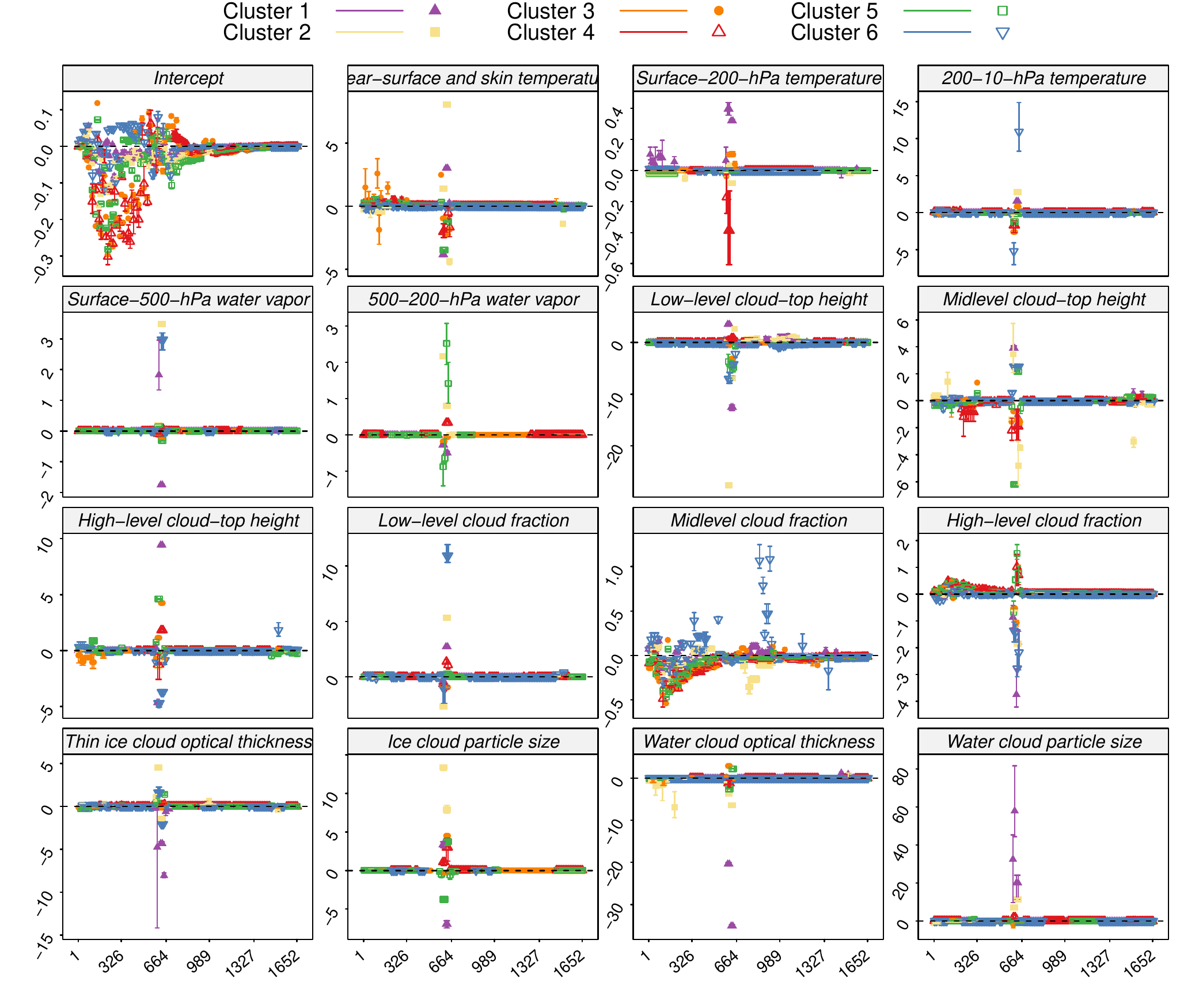}
\vspace{-.9cm}
\caption{ \label{fig:B} 
Posterior summary of $\beta_r$'s under the modal cluster $\hat{\varpi}$ over wavenumber (cm$^{-1}$). 
}
\end{figure}

The clustering $\hat{\varpi}$ in Figure \ref{fig:LY} suggests that the mean levels and variabilities are larger in locations near the equator and have decreasing trends towards to both poles, based on which the global observations are partitioned into $\hat{d} = 6$ clusters. However, the clusters are not rectangles or regular bands that are typical outputs from treed partitions or even Voronoi tessellation on a regular lattice system. 
The map shows that those clusters with large mean levels and variabilities move towards poles during April and November, which further depicts the relationship between those characteristics and temperature. 
The flexibility in producing irregularly-shaped, non-convex clusters by introducing the boundaries with random cluster assignments is helpful in capturing such relevant patterns. 
In addition, 
three clusters near the equator that have relatively higher mean levels of radiances, are still spatially separable since Cluster 4 has less variability with higher homogeneity within the cluster. 
Moreover, the mean level and variability near the north pole are higher than the south pole. 
Moreover, there is one more cluster (Cluster 2) in the southern hemisphere with higher portion of ocean. This is possibly due to that the Antarctic forms a distinct cluster (Cluster 1). 

Next, we report the posterior mean and $95\%$ credible intervals for $\beta_r$ using samples with $\varpi^{(b)} = \hat{\varpi}$ transformed back to the data domain to detect strong signals.
 Figure \ref{fig:B} displays only the significant signals with credible intervals that do not span zero. 
A large portion of $\beta_r$ for most clusters are exactly 0, as suggested by the flatness and vacant wavenumber regions for the functions in Figure \ref{fig:B}. 
It is evident that different covariates can have distinct impacts for the resulting clusters. 
For instance, 
for the temperature group, 
the near-surface and skin temperature has relatively stronger positive impacts at the starting wavenumber regions (approximately from $1$ to $300$ cm$^{-1}$) for Cluster 3, from latitude $50^{\circ}$ S to $20^{\circ}$ S, near the boundary of Tropic of Capricorn and Antarctic Circle. 
While all the 3 temperature factors have strong negative impacts around $660$ cm$^{-1}$ for Cluster 4 near the Equator. The effects become strong, positive for Cluster 1 near the South Pole and Antarctic. 
This is 
reflected by Figure \ref{fig:All}, which shows the surface-200-hPa and 200-10-hPa temperature have varying signs for curves from the two zones, while they all reach the second peak for the control run near that wavenumber.  
For Cluster 6 at the Arctic Ocean, the 200-10-hPa temperature also has strong effects near the valley point where the response radiance has quite small variability and hence strong linearity. 

Another example is the 500-200-hPa water vapor. 
It has strong effects near the second peak point in the control run for Cluster 5, the Tropic of Cancer. 
The cloud-top height and cloud fraction both have remarkably high impacts for multiple latitude zones, though Cluster 6 near the North Pole apparently has relatively high impacts in cloud fraction. 
In particular, the midlevel and high-level cloud fraction correspond to the initial growth of the radiance for the control run. 
This finding echoes with \cite{kato11} in that the cloud fraction change is especially relevant in determining the behaviors of joint and independent effects, such as nonlinearity correspondence and relative difference.  
The last row in Figure \ref{fig:All} shows the thickness and particle size have strong effects for Cluster 1 and Cluster 2, the latitude zones near the South Pole and Antarctic Circle. 
All these facts indicate the importance of our approach of spatial partitioning based on different covariate effects.

\begin{figure}[!ht]
\centering
\includegraphics[scale=.6]{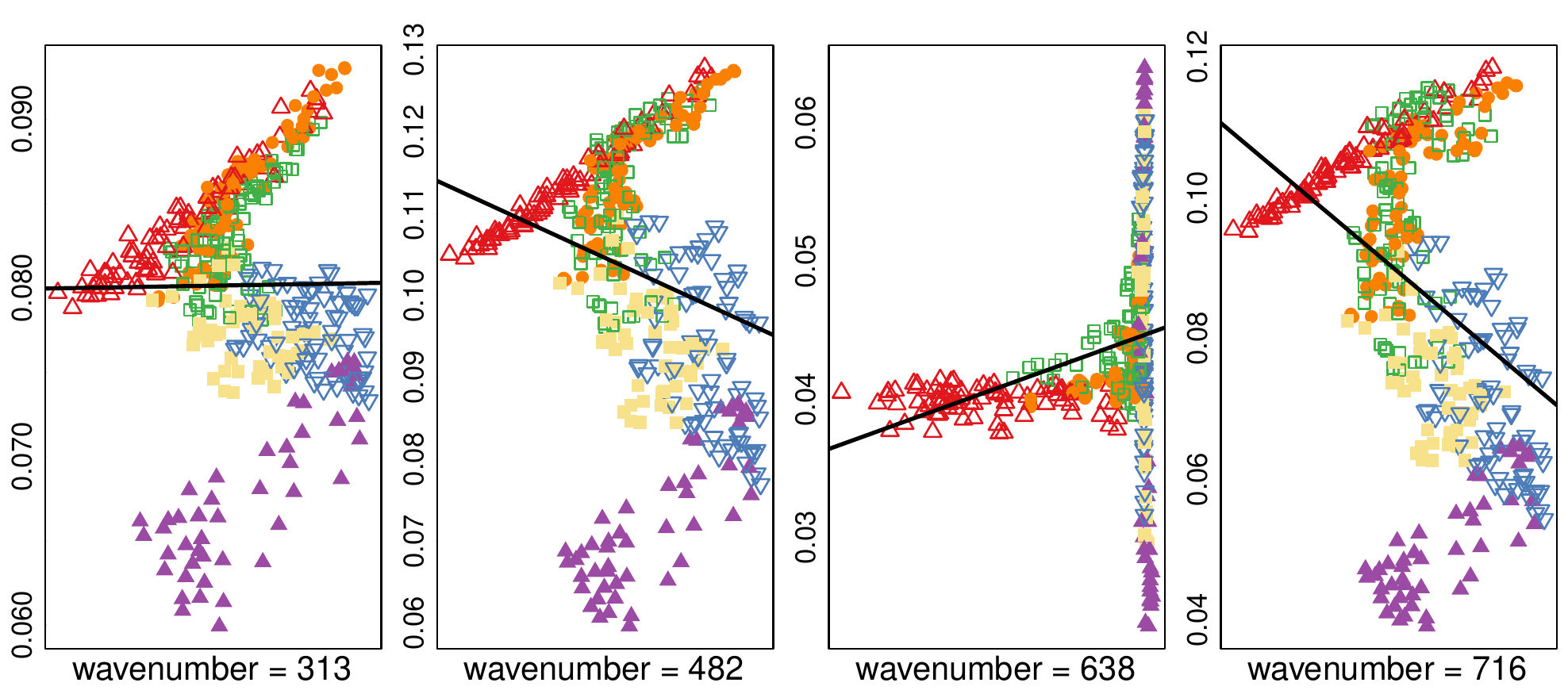}
\vspace{-.21cm}
\caption{ \label{fig:X3} 
An illustration of misfitted regression lines without considering the partition. 
}
\end{figure}

To further validate the estimated fixed-effect functions for capturing the large-scale variation, we also compare the results with the least-square (LS) estimates for each wave number, discarding any individual curve effect, serial and spatial correlation. The LS-estimates in general agree with our estimates in Figure \ref{fig:B}. 
Each panel at a selected wavenumber (cm$^{-1}$) shows $N=432$ pairs of the response radiance versus 200-10-hPa temperature.  
The wavenumber $482$ and $716$ correspond to the two peaks in the response, 
and wavenumber $638$ corresponds to the deep valley where the variability is quite small for the response curves. 
The solid line indicates a local Least-Square estimate. 
The characters and colors are consistently employed in presenting the memberships under the posterior clustering configuration.  
As an example, for wavenumber 313 cm$^{-1}$, the local linear regression may suggest a neutral effect, while the relationships under the partition are significant and varying by groups. 
Furthermore, for 200-10-hPa temperature, the LS-estimates similarly produce a local peak in effect around wavenumber 638 cm$^{-1}$, which corresponds to the deep valley point between the two peak points at wavenumber 482 cm$^{-1}$ and 716 cm$^{-1}$ of the response radiance in Figure \ref{fig:All}. 
Nevertheless, Figure \ref{fig:X3} suggests the local LS-estimates can give quite misleading estimates by discarding the group structure under the posterior clustering configuration from the proposed model. 
With the partition, the proposed model can evidently give better individual estimates of the cluster-specific effect of 200-10-hPa temperature. 
Figure \ref{fig:X3} shows after separation, at the valley point 638 cm$^{-1}$, the near-pole clusters (Cluster 1, 2, and 6) have scatters quite close to an exactly ``vertical'' line. 
This precisely explains the remarkably large effects of 200-10-hPa temperature detected at 638 cm$^{-1}$ in Figure \ref{fig:B} for these clusters. 
It also explains the opposite strong signals in Figure \ref{fig:B} that suggest the effect function of 200-10-hPa temperature does not vary smoothly for Cluster 6: there is a strong negative effect (the ``vertical'' line leans towards the left) detected 
near the strong positive effect  (the ``vertical'' line leans towards the right) at the valley point 638 cm$^{-1}$. 
Similar patterns were found for other covariates, 
which is consistent with the prevalence of strong signal near the valley point 638 cm$^{-1}$ in Figure \ref{fig:B} for almost all the covariates, but for different clusters. 
This finding again demonstrates the importance of identifying regionally varying effects and
providing meaningful partitions that can largely reduce the variability. 

Moreover, Figure \ref{fig:X3} illustrates the merits of spatial contiguity for clusters. 
For instance, the relationship between 200-10-hPa temperature and the response radiance is apparently the same for Cluster 2 and Cluster 6. 
While most clustering algorithms do not incorporate the covariates information, a general clustering algorithm that discards the spatial contiguity may assign the two clusters to one, even when considering the covariate effects. While the two clusters are near different poles and have clearly different interpretations. 
A post-hoc separation may still fail to capture the local features, the boundary information, and the within-cluster spatial dependence.

\section{Discussion}
\label{section:conclusion}

In this work, we present a single Bayesian hierarchical model for answering multiple relevant questions regarding to the spectral radiance study in a unified faction.  
The model has the ability to discover the local impacts in both space and wavenumber regions on the effects of the atmospheric and cloud properties on the spectral radiances from control run. 
The extension to the spatial partitioning models has the enhanced ability to produce clusters with less constraints to capture the seasonal effects, while maintaining the convenience in sampling the partitions to address several commonly recognized issues including the complexity, near-boundary ambiguity, posterior mixing, and uncertainty associated with the grouping structure of the high-dimensional spectral radiances. 
The model can also accommodate the spatial dependence with associated nonstationarity, regional characteristics for wavelet shrinkage. 
Comparing to existing curve clustering methods that can handle spatial data such as 
\cite{rm06}, \cite{konomi2} and \cite{dass}, the proposed SFC model has several attractive features such as producing spatially contiguous clusters, accounting for regional effects of multiple functional covariates, reducing parameters via wavelet shrinkage to alleviate the parameter proliferation issue for product partition models.

From computational perspective, we harness the Bayesian hierarchical model with the two-level parallel computing scheme: between- and within-MCMC parallelism. Due to the developments of high performance computing environments, most independent MCMC runs for Bayesian methods nowadays are conducted in parallel. 
However, the within-MCMC parallelization with random partitions has not been commonly adopted yet partially due to the challenges in preallocation of computing resources for varying sizes and number of components for parallelization. Our exploration in this case study suggests such parallel computing techniques indeed improve the model implementation and the computational ability with partitions that are even random.

The proposed method can work for data with relatively high dimensionality, in particular, when $T> N$ 
(in this application $T=2711$ and $N=432$). 
However, the speedup is not satisfying since the parallel computation in this study mainly takes advantage of the horizontal, rather than vertical partitions of the data. 
The major computational challenge lies in the Gibbs sweep of sampling the $d\times T\times pT$ $(\ugamma, \ubeta)$-parameters. 
Alternatively, 
a sophisticated handle of the high-dimensionality in searching vertical binary partitions can be adopted
in the broad literature of Bayesian variable selection
such as Fast Scan Metropolis-Hastings scheme (FSMH) \citep{br10}, Shotgun Stochastic Search (SSS) \citep{hdw07}, among others. 
In this application, our primary interest is searching for the horizontal partitions in clustering $N$ curves, hence we conduct a vertical reduction in the $T$-functionals that facilitates the search while preserving the shapes.  

The application of the proposed SFC model in this spectral radiance change study suggests it is essential to jointly analyze multiple functional covariates that capture the large-scale variation, while modeling the heterogeneity in covariates effects among subregions in the 2-D input space with latitude zone and month. With full posterior inference on parameters that characterize such regional effects, the full model-based SFC model is more than a clustering model of TOA spectral radiance for providing insight into cloud properties and atmospheric composition. Based on the retrieved feature functions and clustering maps, further investigation and identification of atmospheric composition change in climate study using the TOA spectral radiance change become an immediate task.

\appendix
\label{appendix}

\section{Calculation of the marginal likelihood}
\label{section:appenA}

The full Bayesian hierarchical SFC model proposed through (\ref{prior_v}) to (\ref{prior_e}) includes $3$ different layers of parameters: 
\begin{enumerate}
\item partitioning parameters $\{\varpi, \ugamma\}$, where $\varpi=\{d, \cen_d, L_{\bds}\}$ is horizontal partitioning operator for clustering, and $\ugamma$ is the vertical partitioning operator for variable selection that is nested in $\varpi$ (i.e., it applies to individual cluster);
\item higher-layer parameters $\theta_1=\{ m_{rjk}, \lambda_{rij}, \pi_{rij}, h_{rjk}, \phi_{rjk} \}$; and
\item lower-layer parameters $\theta_2 = \{u_r, \beta_r, \sigma_r^2\}$. 
\end{enumerate}
The partition of parameters into higher and lower layer is due to the tractability of the integrated likelihood for sampling the partitioning parameters.
We further consider integrating the lower-layer parameters $\theta_2$. 
Integrating out the random effect $u_{rjk}$ out from (\ref{lik_r}) with the prior (\ref{prior_u}) involves an integration of the conditional density 
$\pi(u_{rjk} \given \sigma_r^2, \, \beta_r, \, g_{rjk}, \, h_{rjk}, \, \phi_{rjk})$, which is 
$\norm\left(\hat{u}_{rjk},\, \sigma^2_rV_{u,rjk}\right)$ with
\begin{eqnarray}
\label{condi_u}
\left\{
\begin{array}{ccl}
V_{u,rjk} &=& \left(M_{rjk}^{-1} + H_{rjk}^{-1}\right)^{-1}\\
\hat{u}_{rjk} &=& V_{u,rjk}M_{rjk}^{-1}\left( Y_{rjk} - X_{rjk}\beta_r \right).
\end{array}
\right.
\end{eqnarray}
The resulting integrated likelihood, with $\tilde{M}_{rjk} = M_{rjk}+H_{rjk}$, becomes 
\begin{eqnarray}
\label{lik_1}
Y_{rjk} &\indsim& \norm\left(X_{rjk}\beta_r, \,\, \sigma^2_r \tilde{M}_{rjk} \right).
\end{eqnarray}
Next, we continue to integrate out $\beta_r$ from the product of (\ref{lik_1}) over $j,k$ for cluster $\cl_r$, with the prior in (\ref{prior_b}), this involves integrating the conditional density $\pi(\beta_{r} \given \sigma_r^2, \, g_{rjk}, \, h_{rjk}, \, \phi_{rjk})$ which is
$\norm\left(\hat{\beta}_r,\,\, \sigma^2_r V_{\beta,r}\right)$ with
\begin{eqnarray}
\label{condi_b}
\left\{
\begin{array}{ccl}
V_{\beta,r} &=& \left(\Lambda_r^{-1} + \sum_{jk}X_{rjk}'\tilde{M}_{rjk}^{-1}X_{rjk}\right)^{-1}\\
\hat{\beta}_{r} &=& V_{\beta,r}\sum_{jk}X_{rjk}'\tilde{M}_{rjk}^{-1} Y_{rjk}.
\end{array}
\right.
\end{eqnarray}
Let $Y_r=(Y_{r00}', \ldots, Y_{rJ2^{J-1}}')'$, $X_r=(X_{r00}', \ldots, X_{rJ2^{J-1}}')'$, 
$\tilde{M}_r = \mbox{diag}_{jk}\tilde{M}_{rjk}$ and
 $\tilde{\Sigma}_r =\tilde{M}_r + X_{r} \Lambda_rX_{r}'$ for full observations in cluster $\cl_r$. 
As a result, the integrated likelihood with both $u_{rjk}$'s and $\beta_r$ marginalized out, 
involves $Y_{r} \indsim \norm\left(0, \,\, \sigma^2_r \tilde{\Sigma}_{r} \right)$.
Let $\Omega_r = \tilde{\Sigma}_r^{-1}$ be the precision matrix, by Sherman-Morrison-Woodbury formula, we have the closed form $\Omega_r = \tilde{M}_r^{-1} - \tilde{M}_r^{-1}X_r\left(
\Lambda_r^{-1} + X_r'\tilde{M}_r^{-1}X_r
\right)^{-1}X_r'\tilde{M}_r^{-1}$. 
Hence the marginal density can be alternatively viewed as $Y_{r}-X_{r}\hat{\beta}_r\indsim$
\begin{equation}
\label{lik_2}
\norm\left(\zero_{n_r}, \,\, \sigma^2_r 
\left(
\tilde{M}_{r} - X_r\left(
\Lambda_r^{-1} + X_r'\tilde{M}_r^{-1}X_r
\right)^{-1}X_r'
\right) \right),
\end{equation}
due to the fact that $\Omega_r'=\Omega_r$, and
\begin{eqnarray*}
Y_r'\tilde{\Sigma}_r^{-1}Y_r &=& 
Y_r'\Omega_rY_r = Y_r'\Omega_r'\Omega^{-1}_r\Omega_rY_r
= (Y_r - X_r\hat{\beta}_r)'\left( \tilde{M}_r\Omega_r\tilde{M}_r \right)^{-1}
(Y_r - X_r\hat{\beta}_r)\\
&=& (Y_r - X_r\hat{\beta}_r)'\left(
\tilde{M}_{r} - X_r\left(
\Lambda_r^{-1} + X_r'\tilde{M}_r^{-1}X_r
\right)^{-1}X_r'
\right)^{-1}(Y_r - X_r\hat{\beta}_r).
\end{eqnarray*}
We continue to integrate out $\sigma^2_r$ from (\ref{lik_2}) with the Inverse Gamma prior (\ref{prior_e}) when $b_{\sigma}\neq 0$. This involves an integration of the conditional density $\pi(\sigma_r^2 \given g_{rjk}, \, h_{rjk}, \, \phi_{rjk})$ which is
\begin{equation}
\label{condi_e}
\igamma\left( 
a_{\sigma} + n_rT/2, \,\,\, b_{\sigma} +   Y_r'\tilde{\Sigma}_r^{-1}Y_r /2
\right). 
\end{equation}
Consequently, the marginal likelihood under the partitioning operator $\{\varpi, \ugamma\}$ and parameters at higher level $\theta_1=\{ m_{rjk}, \lambda_{rij}, \pi_{rij}, h_{rjk}, \phi_{rjk} \}$, when integrating $\theta_2 = \{u_r, \beta_r, \sigma_r^2\}$ out, is an $n_rT$-dimensional student-$t$ density $Y_{r}-X_r\hat{\beta}_r \given \varpi,  \, \ugamma,  \, \theta_1$ which is
\begin{equation}
\label{lik_3}
\mbox{Student-$t$}\left(\zero_{n_r}, \, \frac{b_{\sigma}}{a_{\sigma}}\left(
\tilde{M}_{r} - X_r\left(
\Lambda_r^{-1} + X_r'\tilde{M}_r^{-1}X_r
\right)^{-1}X_r'
\right), \,
2a_{\sigma}  
\right),
\end{equation}
which clearly depends on the choice of the hyperparameters $(a_{\sigma}, b_{\sigma})$ that capture information of the $\sigma_r^2$, as $b_{\sigma}/a_{\sigma}$ is an estimate of $\sigma_r^2$ that lies in between the prior mode, $b_{\sigma}/(a_{\sigma}+1)$, and the prior mean $b_{\sigma}/(a_{\sigma}-1)$ when $a_{\sigma}>1$. 
It is possible to obtain some accurate prior estimate for experimental data or simulation model output when the nugget effect is trivial. 
One can employ the Jefferys prior $a_{\sigma}=b_{\sigma}=0$ to avoid the sensitivity issue. 
The integrated likelihood then becomes 
\begin{equation}
\label{lik_4}
\pi\left( Y_{r} \given \varpi, \,\ugamma, \,\theta_1 \right) =
|\tilde{\Sigma}_r|^{-1/2}
\left(\pi Y_r'\tilde{\Sigma}_r^{-1}Y_r\right)^{-n_rT/2}
\Gamma\left(n_rT/2\right), 
\end{equation}
The quadratic term $Y_r'\tilde{\Sigma}_r^{-1}Y_r$ and the determinant $|\tilde{\Sigma}_r|$ in (\ref{lik_3}) or (\ref{lik_4}) can be evaluated as
\begin{eqnarray*}
\notag Y_r'\tilde{\Sigma}_r^{-1}Y_r  &=&  Y_r'\Omega_rY_r =
Y_r'\tilde{M}_r^{-1}Y_r - Y_r'\tilde{M}_r^{-1}X_r\left(
\Lambda_r^{-1} + X_r'\tilde{M}_r^{-1}X_r
\right)^{-1}X_r'\tilde{M}_r^{-1}Y_r\\
\label{ml_inv}
&=&Y_r'\tilde{M}_r^{-1}Y_r - \hat{\beta}_r'V_{\beta,r}^{-1}\hat{\beta}_r, \\
\label{ml_det} 
|\tilde{\Sigma}_r| &=& |\tilde{M}_r + X_r\Lambda_rX_r'|
= |\tilde{M}_r| |\Lambda_r^{-1} + X_r'\tilde{M}_r^{-1}X_r| |\Lambda_r|
= |\tilde{M}_r|  |\Lambda_r|/|V_{\beta,r}|,
\end{eqnarray*}
with the latter obtained by Sylvester’s determinant theorem. 
The marginal likelihood (\ref{lik_4}) plays a crucial role in sampling a new partitioning operator $\{\varpi, \gamma\}$. 
It is possible to further integrate out the high-layer parameters $\theta_1$ \citep[see, e.g.,][]{kmh05}, which may however result in densities without convenient forms and hence significantly slow down the stochastic search of the partitions.

\section{MCMC Sampling Procedure} 
\label{section:appenB}

We update the 3-layer parameters as described in Appendix \ref{section:appenA}, 
using a hybrid MCMC scheme that includes a reversible jump step \citep{g95} for sampling the partitions, and subsequent Gibbs steps for sampling the associated parameters.  

\noindent{\bf (1) Update the partitioning operator $\varpi$:} 
Sampling a new clustering configuration $\varpi^*=\{d^*, \cen_{d^*}^*, L_{\bds}^*\}$ 
involves first proposing a ``neighboring'' state that is akin to the current configuration $\varpi=\{d, \cen_d, L_{\bds}\}$ in the model space. 
Let $f(\cdot)$ denote the corresponding proposal mass function. 
Next, we propose the higher-layer parameters $\theta_1=\{ m_{rjk}, \lambda_{rij}, \pi_{rij}, h_{rjk}, \phi_{rjk} \}$ from certain density functions $q(\cdot)$, 
and then propose $\gamma^*$ from the prior given newly proposed $\pi_{rij}^*$. 
Note this step is only required when a new cluster is introduced by $f(\cdot)$ in so-called 
\textit{split} move.   
Finally, the lower-layer parameters $\theta_2$ are subsequently drawn from the full conditional densities (\ref{condi_u}), (\ref{condi_b}) and (\ref{condi_e}). 
The acceptance probability for the proposal
is 
\begin{eqnarray}
\label{ratio_a}
\notag
P_{\mbox{Accept}} &=&\min\left\{1,\quad
\frac{\pi(Y\given \varpi^*, \ugamma^*, \theta_1^*)}{\pi(Y\given \varpi, \ugamma, \theta_1)} 
\,\times\,
\frac{\pi(\theta_1^*\given d^*)\,\pi(L_{\bds^*}\given \cen^*_{d^*})\,\pi(\cen_{d^*}^*\given d^*)\,\pi(d^*)}
{\pi(\theta_1\given d)\,\pi(L_{\bds}\given \cen_d)\,\pi(\cen_d\given d)\,\pi(d)}
\right. \\
 & &\qquad\qquad\qquad\qquad\left. \times\,\,\,
\frac{q(\theta_1\given d, \theta_1^*)\,f(L_{\bds}\given \cen_d)\,f(\cen_d \given \cen_d^*)} {q(\theta_1^*\given d^*, \theta_1)\,f(L_{\bds^*}\given \cen_{d^*}^*)\, f(\cen_{d^*}^*\given \cen_d)}
\,\times\, |\mathcal{J}_{q, \theta,\theta^*}|
\,\,\,\,\right\},
\end{eqnarray}
which is a product of the likelihood ratio, prior ratio, proposal ratio, and the determinant of the Jacobian matrix when $q(\cdot)$ involves deterministic mappings. 
The prior, proposal densities are usually picked to be rather diffuse 
and the Jacobian determinant can be unity, 
hence the first ratio of the marginal likelihood (\ref{lik_3}) dominates the acceptance probability (\ref{ratio_a}). 
One such choice is to propose the parameters from the noninformative or diffuse prior which yields an identity map $q(\cdot)$. This choice obviously causes a high sensitivity to the prior specification, and a noninformative prior/proposal may result in a quite low likelihood for acceptance. 
We instead consider the informative proposal functions for $\theta_1$ when a new cluster $\cl_{r^*}$ is introduced by $f(\cdot)$ from $d$ to $d^*=d+1$ in a split move, by borrowing information from local regions. 
To be more specific, let $\mathcal{E}$ be the set of labels under $\varpi$ for the member locations that form the new cluster $\cl_{r^*}$ in $\varpi^*$. 
We keep the old parameter information $\theta_{1r}$ for all old clusters $\cl_r$'s, and only need to propose $\theta_{1{r^*}}^*$ for $\cl_{r^*}$. 
We propose the $l$th component of $\theta_{1{r^*}}^*$ as a weighted sum, 
$\theta_{1{r^*}l}^* = \varphi^{-1}\left(V_l+\sum_{r\in \mathcal{E}}w_{r}\varphi(\theta_{1rl})\right)$, 
where $\varphi(\cdot)$ is an invertible function that maps $\theta_{1rl}$ from its support to the whole real line, 
$V_l\sim\norm(0,\vartheta^2)$ is an instrumental variable for matching the dimension for this proposal, 
and $w_{r}$ is a spatial weight. 
For example, we take $w_r$ to be the proportion of member locations in the new cluster $\cl_{r^*}$ that are originally from $\cl_r$ to account for the local information. 
We also take $\varphi(\cdot)$ to be $\varphi(\theta_{1rl}) = \log{\theta_{1rl}}$ for the scale parameters $\{m_{rjk}, \lambda_{rij},  h_{rjk}\}\subset \theta_{1}$, and take the 
generalized logistic function for $\{ \pi_{rij},\phi_{rjk}\} \subset\theta_{1}$ with compact support $(A_{rl}, B_{rl})$, i..e, $\varphi(\theta_{1rl}) = \log\{(\theta_{1rl} - A_{rl})/(B_{rl} - \theta_{1rl} )\}$. 
Taking the spatial dependence parameter $\phi_{rjk}\in (\rho_{rn_{{r^*}}}^{-1}, \rho_{r1}^{-1})$, $l$th component in $\theta_1$ as an example, 
we obtain the proposal $\phi_{{r^*}jk}$ by solving the equation
\begin{eqnarray*}
\label{prop_phi}
\hat{\varphi}_l \equiv \varphi(\phi_{{r^*}jk}) = 
\frac{\phi_{{r^*}jk}  - \rho_{{r^*}n_{{r^*}}}^{-1}}{\rho_{{r^*}1}^{-1} - \phi_{{r^*}jk}}
&=&
e^{V_l}
\prod_{r\in \mathcal{E}}
\left(\frac{\phi_{rjk}  - \rho_{rn_r}^{-1}}{\rho_{r1}^{-1} - \phi_{rjk}}\right)^{w_r}. 
\end{eqnarray*}
Hence $\phi_{{r^*}jk} = (\rho_{{r^*}1}^{-1} -\rho_{{r^*}n_{r^*}}^{-1} )/(1+\hat{\varphi}_l) + \rho_{{r^*}n_{r^*}}^{-1}$. 
The corresponding contribution to the Jacobian term in (\ref{ratio_a}) is $|\partial \phi_{{r^*}jk}/\partial V_l| = |\hat{\varphi}_l|(\rho_{{r^*}1}^{-1} -\rho_{{r^*}n_{r^*}}^{-1} )/(1+\hat{\varphi}_l)^2$
since $\partial \hat{\varphi}_l/\partial V_l = \hat{\varphi}_l$.
The 
same argument applies to $\pi_{rij}\in\theta_{1}$ with $(\rho_{rn_{{r^*}}}^{-1}, \rho_{r1}^{-1})$ replaced by the its support $(0,1)$. 
As a consequence, the nontrivial Jacobian term in (\ref{ratio_a}) comprises the contributions by the deterministic function $\varphi$ of all components in $\theta_1$. 
Each proposal function above introduces a tuning parameter $\vartheta^2$ that controls the deviation from the spatially weighted sum for the proposal: a smaller value of $\vartheta^2$ results in less deviation for the proposed parameters from the locally weighted representatives. 
Hence $\vartheta^2$ can be tuned for enhancing the chance of acceptance.  

The \textit{merge} move is conceptually reverse move of the split move with $\varpi^*$ moving to $\varpi$. A randomly selected existing cluster vanishes with members merging into the remaining clusters according to the resulting $\varpi$, and the acceptance ratio is usually the reciprocal of that for the split move. See \cite{zlm14} for more details. 

The labels for the new boundary can be generally proposed from the prior. 
However, a special move is $d^* = d$, i.e., without creation or deletion of any set of parameters during the model search, the procedure focuses on the boundary search for $d$ spatial subregions. 
With a further layer for boundary ambiguity, an immediate challenge is that a center set $\cen_d$ no longer determines a unique label assignment $L$. On the other hand, multiple center sets $\cen_d$'s may result in the same $L$. Therefore, it may not be sensible to adopt the uniform prior $\pi(\cen_d\given d)\propto 1$ on all possible center sets $\cen_d$'s with $d$ clusters, as it can unexpectedly penalize those $\cen_d$'s with larger boundary sets. 
Specifically, if we make a local move of $\cen_d$ to $\cen^*_d$ which introduces a new boundary set $\bds^*$ with new choice sets $K_s^*$'s, the acceptance probability is the minimum of $1$ and the product of ratios
\begin{equation}
\label{ratio_lab}
\frac{\pi(Y\given L_{\bds^*}, \cen_d^*)}{\pi(Y\given L_{\bds}, \cen_d)} \times
\frac{\pi(L_{\bds^*}\given \cen_d^*)}{\pi(L_{\bds}\given \cen_d)} \times
\frac{f(L_{\bds}\given \cen_d)}{f(L_{\bds^*}\given \cen_d^*)} \times
\frac{\pi(\cen_d^*\given d)}{\pi(\cen_d\given d)} \times
\frac{f(\cen_d \given \cen_d^*)}{f(\cen_d^*\given \cen_d)}
\end{equation}
with a proposal density function $f$. 
We consider independently proposing the labels from $f(L_{\bds^*}\given \cen_d^*) = \prod_{s\in \bds^*}f(\ell_s\given \cen_d^*)$ according to 
\begin{eqnarray*}
f(\ell_s = r\given \cen_d^*) = \left\{\begin{array}{lc}
\pi(\ell_s = r\given Y, \cen_d), &\mbox{if } s\in \bds^c\cap \bds^* \\
1, &\mbox{if } s \in \bds\cap \bds^* \mbox{ and } \ell_s = r \mbox{ under }\cen_d\\
1, &\mbox{if } s\in \bds^{*c}  \mbox{ and }  D(s, g_r^*) = \min_{1\leq m\leq d}D(s, g_m^*)\\
0, & {\rm otherwise}
\end{array}\right.
\end{eqnarray*}
and the posterior density $\pi(\ell_s = r\given Y, \cen_d)$
is equal to the likelihood $\pi(Y\given \ell_s = r, \cen_d)$ since we assign equal prior weights $\pi(\ell_s = r \given \cen_d) = 1/|K_s|$. This will cancel with the third ratio of likelihood ratio in (\ref{ratio_lab}).
Therefore, by plugging $\pi(L_{\bds^*}\given \cen_d^*) = \prod_{s\in \bds^*}1/|K_s^*|$, (\ref{ratio_lab}) is reduced to
\begin{eqnarray}
\label{ratio_lab2}
\frac{\prod_{s\in \bds}|K_s|}{\prod_{s\in \bds^*}|K_s^*|} \times
\frac{\pi(\cen_d^*\given d)}{\pi(\cen_d\given d)} \times
\frac{f(\cen_d \given \cen_d^*)}{f(\cen_d^*\given \cen_d)}
\end{eqnarray}
The third ratio, the proposal density ratio for the move between $\cen_d$ and $\cen_d^*$, is generally negligible. 
If we assume the uniform prior on $\cen_d$, i.e., $\pi(\cen_d\given d)\propto 1$, the second ratio additionally vanishes. 
The acceptance ratio becomes $\prod_{s\in \bds}|K_s|/\prod_{s\in \bds^*}|K_s^*|$, and it clearly penalizes $\cen_d^*$ with larger choice sets which typically follow larger boundary sets. 
Therefore, we instead assume the uniform prior on $(\cen_d, L_\bds)$, i.e., $\pi(\cen_d, L_\bds\given d) \propto 1$. Hence $\pi(\cen_d\given d) \propto \prod_{s\in \bds}|K_s|$, and (\ref{ratio_lab2}) becomes $f(\cen_d \given \cen_d^*)/f(\cen_d^*\given \cen_d)$,
which is close to 1. This gives a high chance for the centers to traverse in the spatial domain. 

Once the clustering configuration is updated, the subsequent Gibbs steps are done in parallel, by the conditional independence given $\varpi$.

\noindent{\bf (2) Update the fixed-effects components $(\beta_r, \gamma_r, \lambda_{rij}, \pi_{rij})$:}
First, we consider jointly updating the fixed-effects $\beta_r$ and the indicator of selection $\gamma_r$. 
We sequentially update each component $(\beta_{ri}(jk), \gamma_{ri}(jk))$ given the remaining 
parameters including $(\beta_{ri}(jk)^c, \gamma_{ri}(jk))^c$ and the data. 
We write $\beta_{ri}(jk)^c$ as the remaining $pT-1$ signals with their contributions $X_{si}(jk)^c$, hence $Y_s = X_{si}(jk)\beta_{ri}(jk) + X_{si}(jk)^c\beta_{ri}(jk)^c+ u_s+\epsilon_s$. 
Letting $Y_s^{\ast} = Y_s - X_{si}(jk)^c\beta_{ri}(jk)^c - u_s$, the location-level model (\ref{model1}) indicates 
$Y_s^{\ast} \sim \norm\left(X_{si}(jk)\beta_{ri}(jk), \,\sigma^2_rM_r\right)$. 
By letting $\left(Y_r^{\ast}, X_{ri}(jk)\right)$ pool the observations $\left(Y_s^{\ast}, X_{si}(jk)\right)$ across all locations in cluster $\cl_r$, we write the cluster-level density 
$Y_r^{\ast} \sim \norm\left(X_{ri}(jk)\beta_{ri}(jk), \,\sigma^2_r \uM_r\right)$ with $\uM_r=I_{n_r}\kronecker M_r$. 
We first sample  $\gamma_{ri}(jk)$ through the conditional density that marginalizes $\beta_{ri}(jk)$ out
\begin{eqnarray*}
\pi(\gamma_{ri}(jk) \given Y_r^{\ast}, \pi_{rij}) =  \int \pi(\gamma_{ri}(jk), \beta_{ri}(jk)\given Y_r^{\ast}, \varpi, \sigma_r^2, M_r, \lambda_{rij}, \pi_{rij})\, d\beta_{ri}(jk)\\ 
\propto \pi(\gamma_{ri}(jk)\given\pi_{rij}) 
\int \pi(Y_r^{\ast} \given \beta_{ri}(jk),\sigma_r^2, M_r ) \times
\pi(\beta_{ri}(jk) \given \gamma_{ri}(jk), \lambda_{rij}) \,
d\beta_{ri}(jk)
\end{eqnarray*}
The integration is the density $\zeta_1$ of the $\norm\left(\zero_{n_rT},\, \sigma_r^2( \uM_r+ \lambda_{rij}X_{ri}(jk)X_{ri}(jk)')\right)$ for $\gamma_{ri}(jk)=1$, 
while for $\gamma_{ri}(jk)=0$, it reduces to the density $\zeta_0$ of the
$\norm\left(\zero_{n_rT},\, \sigma_r^2 \uM_r\right)$. 
Combining the prior part $\pi(\gamma_{ri}(jk)\given\pi_{rij})$, we sample 
$\gamma_{ri}(jk)\sim\mbox{Bernoulli}\, (O_{rijk}/(O_{rijk}+1))$, where the posterior odds for $\gamma_{ri}(jk)$ being $1$ is $O_{rijk} = \pi_{rij}\zeta_1/((1-\pi_{rij})\zeta_0)$. 

Next, we sample $\beta_{ri}(jk)$ from the full conditional distribution given the updated $\gamma_{ri}(jk)$. 
If $\gamma_{ri}(jk)=0$, we set $\beta_{ri}(jk)=0$; Otherwise, we sample $\beta_{ri}(jk)$ through
\begin{equation*}
\pi(\beta_{ri}(jk) \given Y_r^{\ast}, \lambda_{rij}, \gamma_{ri}(jk)=1) \,\propto\,
\pi(Y_r^{\ast} \given \beta_{ri}(jk),\sigma_r^2, M_r ) \times
\pi(\beta_{ri}(jk) \given \lambda_{rij}, \gamma_{ri}(jk)=1)
\end{equation*}
which is a Normal distribution with variance $\upsilon^2 = \sigma_r^{2}/(\lambda_{rij}^{-1} + X_{ri}(jk)'\uM_r^{-1}X_{ri}(jk))$ and mean 
$\mu = X_{ri}(jk)'\uM_r^{-1}Y_r^{\ast}/(\lambda_{rij}^{-1} + X_{ri}(jk)'\uM_r^{-1}X_{ri}(jk))$. 
Note that the Bayes factor for calculating posterior odds $O_{rijk}$ for $\gamma_{ri}(jk)$ has the simple expression with $\mu$ and $\upsilon^2$  
\begin{eqnarray*}
\log(\zeta_1/\zeta_0) &=& -\frac{1}{2}\log\left(1+\lambda_{rij}X_{ri}(jk)'\uM_r^{-1}X_{ri}(jk)\right) + \frac{\mu^2}{2\upsilon^2}
\end{eqnarray*}
The full conditional distribution of the signal-to-noise ratio $\lambda_{rij}$ is 
\begin{eqnarray*}
\lambda_{rij}\given\,\beta_r, \gamma_r, \sigma_r^2 &\sim& \igamma(a_{\lambda,ij}^*,b_{\lambda,ij}^*)
\end{eqnarray*}
where $a_{\lambda,ij}^* = a_{\lambda,ij} + \sum_k \gamma_{ri}(jk)/2$ and 
$b_{\lambda,ij}^* = b_{\lambda,ij} + \sum_k\beta_{ri}(jk)^2/(2\sigma_r^2)$.

The full conditional density of the shrinkage probability $\pi_{rij}$ is 
\begin{eqnarray*}
\pi_{rij}|\ugamma &\sim& \mbox{Beta}\left(a_{\pi,ij}+ \sum_{k=0}^{2^j-1}\gamma_{ri}(jk),\,\, b_{\pi,ij}+\sum_{k=0}^{2^j-1}(1-\gamma_{ri}(jk))\right)
\end{eqnarray*}
for  $i=1,\cdots,p, j = 1,\cdots,J$.

\noindent{\bf (3) Update the random-effects components $(u_{rjk}, h_{rjk}, \phi_{rjk})$:}
The full conditional distribution of the scaling parameter $u_{rjk}$ is given in (\ref{condi_u}). 

The full conditional distribution of the scaling parameter $h_{rjk}$ is 
\begin{eqnarray*}
h_{rjk}\given\,u_{rjk}, \sigma_r^2 &\sim& \igamma(a_{h,ij}^*,b_{h,ij}^*)
\end{eqnarray*}
where $a_{h,jk}^* = a_{h,jk} + n_r/2$ and 
$b_{h,jk}^* = b_{h,jk} + u_{rjk}'(F_r - \phi_{rjk}Q_r)u_{rjk}/(2\sigma_r^2)$.

The full conditional density of the spatial dependence parameter $\phi_{rjk}$ is 
\begin{eqnarray*}
\pi(\phi_{rjk}\given\,u_{rjk}, \sigma_r^2) &\propto&
 \exp\{ \phi_{rjk}u_{rjk}'Q_ru_{rjk}/(2h_{rjk}\sigma_r^2)\} \times
\mbox{I}(\phi_{rjk} \in (1/\rho_{rn_r}, 1/\rho_{r1})).
\end{eqnarray*}

\noindent{\bf (4) Update the error components $(\sigma^2_r, m_{rjk})$:}
Let $\epsilon_{rjk} = Y_{rjk} - X_{rjk}\beta_{r} - u_{rjk}$.
The full conditional distribution of the scaling parameter $m_{rjk}$ is 
\begin{eqnarray*}
m_{rjk}\given\,u_{rjk}, \sigma_r^2 &\sim& \igamma(a_{m,rij}^*,m_{h,rij}^*)
\end{eqnarray*}
where $a_{m,rjk}^* = a_{m,jk} + n_r/2$ and 
$b_{m,rjk}^* = b_{m,jk} + \epsilon_{rjk}'\epsilon_{rjk}/(2\sigma_r^2)$.

The conditional distribution of residual variance $\sigma^2$ is 
\begin{eqnarray*}
\sigma^2_r\given \beta_r &\sim& \igamma(a_{\sigma,r}^*, \, b_{\sigma,r}^*)
\end{eqnarray*}
where $a_{\sigma,r}^* = a_{\sigma} + \sum_{ijk}\gamma_{ri}(jk)/2 + n_rT$ and \\
$b_{\sigma,r}^* = b_{\sigma}  + 
\beta_r'\Lambda_r^{-1}\beta_r/2 + \sum_{jk}u_{rjk}'(F_r - \phi_{rjk}Q_r)u_{rjk}/2 + \sum_{jk}\epsilon_{rjk}'\epsilon_{rjk}/(2m_{rjk})$.


\bibliographystyle{jasa} 
\bibliography{Manuscript}

\end{document}